\documentstyle[aps,eqsecnum,epsf]{revtex}
\begin{document}
\voffset=0.5truein
\draft

\wideabs{
\title{
Gravitational Waves from a Compact Star in a Circular, Inspiral Orbit,
\\ in the Equatorial Plane of a Massive, Spinning Black Hole, as Observed by
LISA 
}
\author{Lee Samuel Finn,$^{(1)}$ and Kip S.  Thorne$^{(2)}$
}
\address{
$^{(1)}$Department of Physics, Astronomy \& Astrophysics, The Pennsylvania
State University, University Park, PA 16802 \\
$^{(2)}$Theoretical Astrophysics, California Institute of Technology, Pasadena, CA 91125 \\
}
\date{Received 7 April 2000}
\maketitle
\begin{abstract}

Results are presented from high-precision computations of the orbital
evolution and emitted gravitational waves for a stellar-mass object
spiraling into a massive black hole in a slowly shrinking, circular,
equatorial orbit.  The focus of these computations is inspiral near the
innermost stable circular orbit (isco)---more particularly, on orbits
for which the angular velocity $\Omega$ is $0.03 \alt \Omega/\Omega_{\rm isco}
\le 1.0$.  The computations are based on the Teuksolsky-Sasaki-Nakamura
formalism, and the results are tabulated in a set of functions that are
of order unity and represent relativistic corrections to 
low-orbital-velocity formulas.  These tables can form a foundation for
future design studies for the LISA space-based
gravitational-wave mission.  A first survey of applications to LISA
is presented:  Signal to noise ratios $S/N$ are computed and graphed as 
functions of the time-evolving gravitational-wave frequency for the
lowest three harmonics of the orbital period, and for various
representative values of the hole's mass $M$ and spin $a$ and the
inspiraling object's mass $\mu$, with the distance to Earth chosen to be 
$r_o = 1$ Gpc.  These $S/N$'s show a very strong dependence
on the black-hole spin, as well as on $M$ and $\mu$.  Graphs are
presented showing the range of
the $\{M,a,\mu\}$ parameter space, for which $S/N > 10$ at $r_o =
1$ Gpc during the last year of inspiral. 
The hole's spin $a$ has a factor $\sim 10$ influence
on the range of $M$ (at fixed $\mu$)
for which $S/N>10$, and the presence or absence of a
white-dwarf-binary background has a factor $\sim 3$ influence.  A
comparison with predicted event rates shows strong promise for detecting
these waves, but not beyond about 1Gpc if the inspiraling object is a
white dwarf or neutron star.  This argues for a modest lowering of
LISA's noise floor.  A brief discussion is given of the prospects for 
extracting information from the observed waves.

\end{abstract}
}
\narrowtext

\section{INTRODUCTION AND SUMMARY}
\label{sec:Introduction}

Earth-based gravitational-wave detectors operate in the high-frequency
band, $\sim 1$---$10^4$ Hz, in which lie the waves from black holes of
masses $\sim 2$---$10^3 M_\odot$.  Space-based detectors operate in
the low-frequency band, $\sim 10^{-4}$--$1$ Hz populated by waves from
black holes of mass $\sim 10^3$--$10^8 M_\odot$.  The high-frequency band
is likely to be opened up early in the next decade by the LIGO-VIRGO network 
of earth-based detectors \cite{ligo_virgo}.  The premier instrument for the 
low-frequency band is the Laser Interferometer Space Antenna (LISA)
\cite{lisa}.

The European Space Agency has selected LISA 
as one of three ``Cornerstone'' missions in its ``Horizon 2000+'' program,
NASA has appointed a mission definition team for LISA, and the ESA and NASA
teams are talking to each other informally about the possibility of flying
LISA as a joint ESA/NASA mission in the $\sim 2010$ time frame.

One of the most interesting and promising gravitational wave sources for LISA
is the final epoch of inspiral of a compact, stellar-mass object into a
massive black hole.  In the LISA frequency band, where the central hole 
must have
$M\alt 10^8 M_\odot$, all giant stars and 
main-sequence stars 
will be tidally disrupted before the end of their inspiral, 
but compact objects---white dwarfs, neutron
stars, and small black holes---can survive intact.  (Depending on 
the hole's spin, a massive white dwarf will be disrupted before
the end of inspiral if $M<M_{\rm max} \sim 10^4$--$10^5 M_\odot$.
Neutron stars and small black holes can never be tidally disrupted 
in the LISA frequency band.)

Sigurdsson and Rees \cite{sigurdsson_rees} have estimated the event rate
for such compact objects to spiral into massive black 
holes. ``Assuming most spiral galaxies have a
central black hole of modest mass ($\sim 10^6 M_\odot$) and a cuspy spheroid,''
and for ``very conservative estimates of the black hole masses and central
galactic densities,'' they estimate one inspiral 
per year within 1Gpc distance of Earth.  Most of the inspiraling objects are
likely to be white dwarfs or neutron stars; the inspiral rate for  
stellar-mass black holes ($\mu \sim 6$--$10M_\odot$) may be
ten times smaller, about 3 per year out to 3Gpc, according to Sigurdsson
\cite{sigurdsson}. Sigurdsson notes, however, that 
that the evidence for a recent burst of star formation 
in the central region of our
galaxy suggests that normal nucleated spirals might have such starbursts 
every
$\sim 10^8$ years, which would enhance the stellar-mass black-hole density
by a factor $\sim 10$ and would lead to stellar-mass black-hole inspirals
of one per year out to one Gpc.  He notes, further, that if there was just 
one $50M_\odot$ black hole in the core of each galaxy now containing a $\sim
10^6M_\odot$ central black hole, the result would be several inspirals of
such $50M_\odot$ holes per year out to a cosmological redshift $z=1$,
all readily observable by LISA. 

LISA's observations of waves from
such inspirals will have major scientific payoffs
\cite{phinney}:
\begin{description}
\item[$\bullet$] Ryan \cite{ryan1} has shown that for circular equatorial
orbits, the waves will carry, 
encoded in themselves, a map of the vacuum spacetime metric of the central 
black hole (or, equivalently, the values of the hole's multiple
moments), and he
has made a first, very crude, estimate of the precision with which LISA can
extract that map \cite{ryan2}. Ryan's estimate is quite promising.  From
the extracted map, one can determine whether the hole's geometry is
that of the Kerr metric (i.e. ``test the black-hole no hair theorem''),
and one can use such maps to
search for other kinds of conjectured massive central bodies 
(e.g. soliton stars \cite{soliton_star} and naked singularities).
It seems likely that this is true not only for circular geodesic orbits,
but also for generic orbits.
\item[$\bullet$] 
The observation of many such events will provide (i) a census of the masses
and spins of the massive central holes, (ii) a census of the masses of the 
inspiraling objects (which depend on and thus tell us about the 
initial stellar mass function and mass
segregation in the central parsec of galactic nuclei), and (iii) a census of
event rates (which depend on physical processes and on gravitational
potentials in the central parsec). 
\item[$\bullet$] 
In active galactic nuclei, the inspiral orbit may be significantly
affected by drag in an accretion disk, producing both complications in
the interpretation of the observations and opportunities for learning about
the disks' mass distribution \cite{chakrabarti}.
\end{description}

In planning for the LISA mission, it is important to understand the details of
the waves emitted by such inspirals.  Those details are
the most important factors in the choice of the mission's noise floor and its
duration, and are likely to be the principal drivers of its data
analysis requirements and algorithms.

The foundations for computing the emitted waves are nearly all in place:
\begin{description}
\item[$\bullet$]
If the orbit is known, then the waveforms and strengths can be computed
using the Teukolsky \cite{teukolsky} -- Sasaki-Nakamura
\cite{sasaki_nakamura} (TSN) formalism for first-order perturbations of Kerr
black holes.  
\item[$\bullet$]
The orbital evolution is governed by radiation reaction
(and, if there is a robust accretion disk present, by accretion-disk
drag \cite{chakrabarti}).  Most massive holes are in galaxies with
normal (non-active) nuclei, and are thought to be surrounded by tenuous disks
with ``advection-dominated accretion flow'' (ADAF).  Narayan \cite{narayan} has
shown that accretion drag should be totally negligible in such ADAF disks, so
the orbital evolution is very cleanly governed by radiation reaction.  This is
the situation that we analyze in this paper; we ignore accretion-disk drag.
Those few holes that are in active galactic nuclei may be surrounded by
``thin'' or ``slim'' accretion disks, for which Chakrabarti and colleagues
\cite{chakrabarti} have shown that accretion-disk drag may be significant.

\item[$\bullet$] 
The radiation reaction's influence on the orbit can be characterized fully 
by the rates of change of three ``constants'' of the orbital motion: the
orbital energy $E$, axial component of angular momentum $L$, and Carter
constant $Q$ \cite{carter}.  From the emitted waves (computed via
the TSN formalism), one can read off $\dot E \equiv dE/dt$ and
$\dot L$; but the only known way to compute $\dot Q$ is directly
from the radiation reaction force.
\item[$\bullet$] A formal expression for the radiation reaction force
has been derived recently by Mino et.~al.\ \cite{mino} and by Quinn and Wald
\cite{quinn_wald}, and several researchers are now working hard to
convert this into a practical computational tool for deducing $\dot Q$
\cite{rrcurrent}.
This will complete the necessary set of tools for computing all details
of the emitted waves.
\end{description}

The emitted waves will be so complex and so rich in structure and
in parameter dependence, that it will require extensive computations
to give us the full knowledge required by the LISA mission.  Those
computations are proceeding in stages:
\begin{enumerate}
\item Initial quick surveys, based on the Newtonian or quasi-Newtonian
orbits and the quadrupole-moment approximation to gravitational-wave
emission.  Such surveys are the foundation
for the event rate estimates by Sigurdsson and Rees discussed above. 
\item More detailed and accurate surveys for orbits in the massive hole's 
equatorial plane, using the TSN formalism.  Such
surveys do not require  computing $\dot Q$, since $Q$ vanishes for
equatorial orbits.  These surveys are of several types:
\begin{enumerate}
\item Studies of the evolution of the orbit's eccentricity.  Such
studies have been carried out by Tanaka et.~al.\ \cite{tanaka,mino_review} 
and Cutler et.~al.\ \cite{cutler_kennefick} for non-spinning holes, and by 
Kennefick \cite{kennefick} for small eccentricities around spinning holes.  
These studies, coupled with estimates of the orbital eccentricities 
when the objects are far from the hole and are being frequently perturbed by 
near encounters with other objects \cite{hils_bender,sigurdsson_rees}, suggest 
that, despite the circularizing effect of radiation reaction, the  
eccentricities will still typically be large, 
$e \agt 0.3$, when the object nears the hole's horizon. 
\item Systematic computations of the details of the emitted waves and
the orbital evolution for circular, equatorial orbits.  This paper presents 
such computations and a
companion paper \cite{ori_thorne} extends them to the {\it transition
regime}, near the innermost stable circular orbit (isco), during which the
orbit makes a gradual transition from adiabatic inspiral to a plunge into the
hole.
\item Computations of the waves' details and orbital evolution
for elliptic, equatorial
orbits.  First explorations have been carried out by Shibata 
for general ellipticity \cite{shibata94}
and by Kennefick for small ellipticity 
\cite{kennefick} (though in the 1970s and 1980s there were studies for
equatorial orbits that plunge from radial infinity into a hole or scatter off
a hole \cite{PlungeScatter}.
\end{enumerate} 
\item Surveys of the orbital evolution and 
waves for circular orbits out of the hole's equatorial plane.  It is known 
that radiation reaction drives circular orbits into circular orbits, 
thereby causing $\dot Q$ to
evolve in a manner that is fully determined by TSN-formalism
calculations of $\dot E$ and $\dot Q$ \cite{kennefick_ori,ryan3,mino_thesis}.
Therefore, the tools
are fully in hand for these surveys, and Hughes \cite{hughes} is in the late
stages of the first one.  (See Shibata \cite{shibata93} for an 
exploration of the wave emission before anyone knew how, correctly, to compute
the orbital evolution, and see Shibata et.\ al.\ 
\cite{shibataetal95} for studies of
orbits with very small inclination angles to the equatorial plane.)
\item Surveys of orbital evolution and waves for the generic, most
realistic situation: elliptic orbits outside the equatorial plane.  Such
surveys must await a practical computation technique for $\dot Q$.
\end{enumerate}

For circular, equatorial orbits (the subject of this paper), there have been
extensive previous calculations, beginning with the pioneering study
by Detweiler \cite{detweiler}; for a review see Mino et.~al.\ 
\cite{mino_review}.  However, these previous calculations have been motivated
by the needs of LIGO/VIRGO observations in the high-frequency band,
where (i) the ratio $\mu/M$ of object mass to hole mass is not
very small, so finite-mass-ratio effects (omitted by the TSN formalism)
are important, and 
(ii) almost all of the observed inspiral signal comes from radii
large compared to the hole's horizon, so post-Newtonian techniques can
be used.  The previous calculations have focused almost entirely on
carrying the post-Newtonian calculations to very high order, on
developing techniques for accelerating their convergence, and---via
comparison with TSN calculations---on evaluating their convergence
\cite{mino_review}. 

LISA's regime and needs are quite different from this.  For LISA, most of the
signals are likely to come from systems with extreme mass ratios, $\mu
/M \ll 1$, for which (a) the TSN formalism is highly accurate and (b) the
object lingers for a very long time in the
vicinity of the hole's horizon before plunging into it.  This
means that post-Newtonian calculations are neither needed, nor
appropriate.

Because of these differences between the LIGO-VIRGO regime and the LISA
regime, the previous TSN-based studies do not serve LISA's needs.  The
purpose of this paper is to begin filling that gap.  Specifically:

In this paper we introduce a new set of functions 
$\cal N$, $\cal T$, $\dot{\cal E}$, to characterize the orbital evolution
and the emitted waves.  These functions are dimensionless and of order unity,
and depend on the hole's dimensionless spin parameter $a =
($angular momentum$)/M^2$ and on the orbit's dimensionless radius
$\tilde r = r/M$.  We give extensive tables of these
functions, as computed by one of us (LSF) using the TSN formalism.  We
then use those tables to compute the evolution of the waves' frequency
and signal strength in LISA for a number of instructive values
of the parameters $M =($hole mass), $a = ($hole spin parameter), $\mu =
($object mass), and $m = ($wave harmonic order$) \equiv ($wave
frequency$)/($orbital frequency).  From these computations we draw a
number of conclusions of importance for the LISA mission.

The paper is organized as follows.
Our notation, including the dimensionless functions ${\cal N}$, ${\cal
T}$, ..., is introduced in Sec.\ \ref{sec:Notation}.  Formulas for
computing the dimensionless functions, and formulas for the orbital
evolution and the waves' properties are given in Sec.\
\ref{sec:Formulas}.  Tables of the dimensionless functions are given
and discussed in Sec.\ \ref{sec:Tables}. 
Applications to LISA are presented in Sec.\ \ref{sec:Applications}.
Finally, concluding remarks are given in Sec.\
\ref{sec:Conclusions}.

\section{Notation}
\label{sec:Notation}

In this paper we shall adopt the following notation to describe
the compact object's inspiral and the gravitational waves it emits;
throughout we use geometrized units, i.e.\ we set $G\equiv($Newton's
gravitation constant$)=1$ and $c\equiv$(speed of light$)=1$. 

\begin{description}

\item[$\mu$:] The mass of the inspiraling object. 

\item[$M$:] The black hole's mass.

\item[$\eta \equiv \mu/M$:] The mass ratio, assumed $\ll 1$.

\item[$a \equiv S/M^2$:] The hole's ``rotation parameter;'' 
here $S$ is the hole's spin angular momentum.

\item[$r$:] The orbit's Boyer-Lindquist radial coordinate; defined by
$\sqrt{r^2 + a^2(1+2M/r)} = (1/2\pi)\times$(the object's 
orbital circumference)

\item[tilde:] A tilde over a quantity means that it has been made
dimensionless by multiplying by the appropriate
power of $M$ and, when the quantity is $\propto \mu$, multiplying by a factor
$1/\mu$. 

\item[$\tilde r \equiv r/M$:] The dimensionless radius of the orbit.

\item[$\Omega$:] The object's orbital angular velocity, as measured using 
Boyer-Lindquist coordinate time $t$ (defined below), i.e., using clocks that
are far from the hole and at rest with respect to it.

\item[$\tilde\Omega \equiv M\Omega$:] The dimensionless orbital 
angular velocity, which is related to $\tilde r$ by $\tilde\Omega =
1/(\tilde r^{3/2}+a)$; cf.\ Eq.\ (2.16) of Ref.\ \cite{bpt}.  
When $\tilde \Omega$
is small (large $\tilde r$), Kepler's laws dictate that $\tilde\Omega
\simeq (M/r)^{3/2} = ($orbital velocity$)^3$, i.e.\ (orbital velocity$) \simeq
\Omega^{1/3}$.

\item[subscript isco:]  A quantity evaluated at the object's innermost
stable circular orbit (``isco''), where the inspiral ends and 
the plunge begins; 
for example, $\tilde\Omega_{\rm isco}$ is the value of $\tilde\Omega$ at the 
isco.

\item[$t$:]  Boyer-Lindquist coordinate time, or equivalently time as
measured at ``radial infinity'' or on Earth.

\item[$T$:]  The Boyer-Lindquist time $\Delta t$ until the isco 
is reached;
i.e., the total remaining duration of the inspiral.

\item[$N_{\rm orb}$:]  The number of orbits remaining until the isco
is reached.

\item[$r_o$:] The distance from the binary to Earth.

\item[$m$:] The order of a harmonic of the orbital frequency. 

\item[$f_m = (m/2\pi)\Omega$:] The frequency of gravitational waves
in the $m$'th harmonic.

\item[$E$:]  The object's total energy including rest mass, i.e., the
component $-p_t$ of its 4-momentum. Note that, because the object is
gravitationally bound to the black hole, $E<\mu$, its gravitational
binding energy is $\mu - E >0$.

\item[$\tilde E\equiv E/\mu$.]

\item[$\dot E_\infty$:] The total rate of emission of energy into gravitational
waves that go to infinity.

\item[$\dot E_H$:] The total rate of emission of energy into gravitational
waves that go down the horizon.

\item[$\dot E_{\rm GW} \equiv \dot E_\infty+\dot E_H$:] The total rate of 
emission of energy into gravitational waves
that go both to infinity and down the hole's horizon; and also, by
energy conservation, the rate of decrease of the object's total energy; i.e.,
$\dot E_{\rm GW} \equiv dE_{\rm GW}/dt = -dE/dt$.

\item[$\dot E_{\infty m}$:] The total rate of emission of energy into the $m$'th
harmonic of the waves that go to infinity. 

\item[$h_{o,m} \equiv \sqrt{\langle{h_{m+}}^2 + 
{h_{m\times}}^2\rangle}$:]  The {\it rms amplitude} of the
gravitational waves in harmonic $m$ emitted toward infinity, 
at a time when the wave frequency is
$f_m$; here $h_{m+}(t,{\bf n})$ and $h_{m\times}(t, {\bf n})$ are the two
waveforms emitted in a direction $\bf n$ and arriving at the Earth's
distance $r_o$; $\langle \ldots \rangle$
is an average over $\bf n$ and over a period of the waves; and the
average over time automatically produces a factor $1/2$ thereby making
$h_{o,m}^2$ be 
the mean value of ${1\over2}[($amplitude of $h_{m+})^2 + ($amplitude of
$h_{m,\times})^2]$.

\item[$h_{c,m} \equiv h_{o,m}\sqrt{2f_m^2/\dot f_m}$]:  A {\it
characteristic amplitude} for the waves in harmonic $m$; here and
throughout this paper the dot denotes
a time derivative. The significance of $h_{c,m}$ is discussed below. 

\item[$h'_{c,m}$] $\equiv h_{c,m} \min[1,\sqrt{3(1-f_m/f_{m,{\rm isco}})}]$:
A modified characteristic amplitude, discussed below.  

\item[$h_n(f) \equiv \sqrt{f S_h^{\rm SA}(f)}$]: 
LISA's ``sky-averaged'' rms noise
in a  bandwidth equal to frequency $f$. 
Here $S_h^{\rm SA}(f)$ is the one-sided spectral density $S_{h_+}(f)$ for some
linear polarization $+$, inverse averaged over source directions and
polarization (``sky-averaged''), $1/S_h^{\rm SA} \equiv 
\langle1/S_{h_+}\rangle$.

\end{description}

From the general relation $dE_{\rm GW}/dtdA = (1/16\pi)({\dot h_+}^2 +
{\dot h_\times}^2)$ for the energy flux in gravitational waves in terms of
the time derivatives of the two waveforms (Eq. (10) of
Ref.~\cite{300yrs}), we infer that the 
rms amplitude and the energy in harmonic $m$ are related to each other
by
\begin{equation}
h_{o,m} = {2\sqrt{\dot E_{\infty m}} \over m\Omega r_o}\;.
\label{ampenergy}
\end{equation}

We shall use, as our measure of where the object is in its orbit,
the dimensionless orbital angular frequency 
$\tilde \Omega$, which is related to the gravitational-wave frequency in
harmonic $m$ by $f_m =(m/2\pi M) \tilde\Omega$.  We shall write
various fully relativistic, time-evolving quantities ($\dot E_{\rm GW}$,
$h_{c,m}$, etc.) as the leading-order (``Newtonian'') term in an
expansion in $\tilde\Omega^{1/3} \simeq ($orbital velocity),
multiplied by relativistic corrections.  Our notation for
the relativistic corrections will be the following:

\begin{description} 

\item[${\cal N}$:]  The correction to $\Omega^2/\dot\Omega\equiv
\tilde\Omega^2/\dot{\tilde\Omega}$, where the
dot is a time derivative.  Note that $\Omega^2/\dot\Omega =
d\Phi/d\ln\Omega$ is the number of radians $d\Phi$ of orbital motion
required to produce (due to radiation reaction) a fractional change
$d\Omega/\Omega$ in the orbital frequency.

\item[${\cal N}_{\rm orb}$:] The correction to $N_{\rm orb}$ (the number of
orbits remaining until the end of the inspiral).

\item[${\cal T}$:]  The correction to $T$ (the remaining time to the end of
the inspiral).

\item[$\dot {\cal E}$:]  The correction to $\dot E_{\rm GW}$ 
(the total energy loss rate).  

\item[$\dot {\cal E}_{\infty m}$:]  The correction to $\dot E_{\infty m}$ 
(the energy radiated to infinity in harmonic $m$). 

\item[${\cal H}_{o,m}$:]  The correction to $h_{o,m}$ (the rms wave
amplitude in harmonic $m$). 

\item[${\cal H}_{c,m}$:]  The correction to $h_{c,m}$ (the
characteristic amplitude in harmonic $m$).
\end{description}

The {\it characteristic amplitude} $h_{c,m}$ needs some explanation.  
As the object spirals inward in its orbit, its $m$'th harmonic waves spend
$\sim f_m^2/\dot f_m = d\Phi_m/(2\pi d\ln f_m)$ cycles in the vicinity of
frequency $f_m$ (where $\Phi_m$ is the harmonic's phase). 
Correspondingly, in a detector that observes the waves throughout the inspiral
epoch $\Delta t = f_m/\dot f_m$, the signal 
is enhanced, in comparison to the detector noise, by approximately the
square root of this quantity.  The signal strength is thus approximately
the same as
would be produced by a broad-band burst of amplitude 
$h_{c,m} \equiv h_{o,m}\sqrt{2f_m^2/\dot f_m}$. 

The factor $2$ inside the square root arises from a more precise 
definition of $h_{c,m}$ \cite{flanagan_hughes}:
The signal to noise ratio produced by the waves' $m$'th harmonic, averaged over
all possible orientations of the source and the detector, is given by
\begin{equation}
\left({S\over N}\right)_{\rm rms} = \sqrt{\int \left[
{h_{c,m}(f_m)\over h_{\rm n}(f_m)}\right]^2 d\ln f_m}\;,
\label{SoverN}
\end{equation}
where $h_n(f_m)$
is the detector's rms noise at frequency $f_m$, in a bandwidth equal to 
frequency, averaged over the sky.  Equation
(\ref{SoverN}) serves as a definition of $h_{c,m}(f)$.  The relation
$h_{c,m}(f_m) = h_{o,m}(f_m) \sqrt{2f_m^2/\dot f_m}$ then follows from Eq.\
(29) of Ref.\ \cite{300yrs} (with the factor $2$ changed to $4$ to correct an
error), together with the definition of $h_{o,m}$ given above, Eqs.\
(\ref{SoverN}) and (\ref{ampenergy}), 
and the evaluation of Fourier transforms using
the stationary phase approximation.

When the inspiraling object nears the isco, the bandwidth available for
building up its signal in the detector becomes 
less than $\Delta f = f$.  A good measure of this reduced bandwidth is
$\Delta f = 2(f - f_{\rm isco})$ 
(with half of this band below $f$ and half above).
This is less than $f$ for $2f_{\rm isco}/3<f<f_{\rm isco}$.
Correspondingly, the amplitude of the built-up signal is $\sim h_c
\sqrt{2(f_{\rm isco}-f)/(2f_{\rm isco}/3)}$.  Our modified characteristic
amplitude  
$h'_{c,m} \equiv h_{c,m} \min[1,\sqrt{3(1-f_m/f_{m,{\rm isco}})}]$
takes this signal reduction into account.

\section{Formulas for Inspiral and Waves}
\label{sec:Formulas}

In this section we shall give leading-order (in $\tilde\Omega^{1/3}$)
formulas for the various
time-evolving quantities, as functions of the dimensionless
orbital angular frequency $\tilde\Omega$ and black-hole spin $a$, 
and thereby we shall produce
exact definitions of the relativistic correction functions.  To make
clear the magnitudes of various quantities, we shall write some of our
formulas numerically in a form relevant 
to LISA (for which we choose as a fiducial frequency $f_2=0.01{\rm
Hz}$ and as a fiducial source, a $\mu=10 M_\odot$ black
hole spiraling into a $M=10^6 M_\odot$ hole at $r_o=1$ Gpc distance 
from Earth).  We shall also write our formulas
in a form relevant to the LIGO-VIRGO network
of high-frequency detectors (with, as our
fiducial frequency, $100{\rm Hz}$, and our 
fiducial source, a $1 M_\odot$ neutron star spiraling into a $100 M_\odot$
hole at 1Gpc distance).

In the Newtonian limit, the orbital radius and orbital angular 
velocity are
linked by the Keplerian relation
\begin{equation}
{M\over r} \equiv {1\over \tilde r} = (M\Omega)^{2/3} \equiv \tilde\Omega^{2/3}.
\label{Omega_r}
\end{equation}
This permits us to write the number of orbital radians spent near
orbital angular frequency $\Omega$ in the following form
(cf.~Eqs.~(3.16) of MTW \cite{mtw}, in which $a$ is our orbital radius
$r$):
\begin{eqnarray}
{\Omega^2\over\dot\Omega} &=& {d\Phi\over d\ln\Omega}=
{5\over96}{1\over\eta}{1\over\tilde\Omega^{5/3}}{\cal N} \nonumber\\
&=& {1.17\times10^5\over(f_2/.01{\rm Hz})^{5/3}}
\left({10M_\odot\over\mu}\right)
\left({10^6M_\odot\over
M}\right)^{2\over3}{\cal N} 
\label{omega2overdotomega} \\
&=& {117\over(f_2/100{\rm Hz})^{5/3}}
\left({1M_\odot\over\mu}\right)
\left({100M_\odot\over
M}\right)^{2\over3}{\cal N}\;. \nonumber
\end{eqnarray}
Here $\cal N$ is the general relativistic correction, which is unity in
the ``Newtonian'' limit $\tilde\Omega\ll1$.  
Similarly, the total remaining time until the end
of the inspiral (Eq.~(36.17b) of MTW) is
\begin{eqnarray}
T &=& {5\over 256}{1\over\eta}{M\over\tilde\Omega^{8/3}}{\cal T}\nonumber\\
&=&{1.41\times10^6{\rm sec}\over(f_2/.01{\rm Hz})^{8/3}}
\left({10M_\odot\over\mu}\right)
\left({10^6M_\odot \over M}\right)^{2\over3}{\cal T}\label{T}\\
&=&{0.141{\rm sec}\over(f_2/100{\rm Hz})^{8/3}}
\left({1M_\odot\over\mu}\right)
\left({100M_\odot\over
M}\right)^{2\over3}{\cal T}\;,\nonumber
\end{eqnarray}
and the number of orbits remaining until the end of the inspiral is
\begin{eqnarray}
N_{\rm orb} &=& {1\over2\pi}\int_{\ln\Omega}^{\ln\Omega_{\rm isco}} {d\Phi\over
d\ln\Omega} d\ln\Omega = {1\over64\pi}{1\over\eta}
{1\over\tilde\Omega^{5/3}}{\cal N}_{\rm orb}\nonumber\\
&=& {1.11 \times 10^4\over(f_2/.01{\rm Hz})^{5/3}} 
\left({10M_\odot\over\mu}\right)
\left({10^6M_\odot \over M}\right)^{2\over3}{\cal N}_{\rm orb}\label{Norb}\\
&=&{11.1\over(f_2/100{\rm Hz})^{5/3}}
\left({1M_\odot\over\mu}\right)
\left({100M_\odot\over
M}\right)^{2\over3}{\cal N}_{\rm orb}\;.\nonumber
\end{eqnarray}

By integrating Eq.~(\ref{omega2overdotomega}) inward to the isco,
one can derive the following expression for
the general relativistic corrections 
$\cal T$ for $T$ and ${\cal N}_{\rm orb}$ for $N_{\rm orb}$
in terms of that $\cal N$ for $\Omega^2/\dot\Omega$:
\begin{equation}
{\cal T} = {8\over3}
\tilde\Omega^{8/3}\int_{\tilde\Omega}^{\tilde\Omega_{\rm isco}} {{\cal
N}d\tilde\Omega \over \tilde\Omega^{11/3}},
\label{calT}
\end{equation}
\begin{equation}
{\cal N}_{\rm orb} = {5\over3}
\tilde\Omega^{5/3}\int_{\tilde\Omega}^{\tilde\Omega_{\rm isco}} {{\cal
N}d\tilde\Omega \over \tilde\Omega^{8/3}},
\label{calNorb}
\end{equation}

The total energy loss rate [Eq.~(3.16) of
MTW] is
\begin{equation}
\dot E_{\rm GW}  = - \dot E
= {32\over 5} \eta^2\tilde\Omega^{10/3}\dot {\cal E},
\label{dotE}
\end{equation}
where $\dot {\cal E}$ is the general relativistic correction. 

When the object is at large radii (small $\tilde\Omega$),
the power radiated by the system's mass multipole moments $I^{l,\pm m}$ 
is of order $\eta^2 \tilde\Omega^{2+2l/3}$, while that radiated by its current
multipole moments $S^{l,\pm m}$ is of order $\eta^2\tilde\Omega^{2+2(l+1)/3}$
\cite{rmp}.  Correspondingly, 
the power $\dot E_{\infty m}$ radiated to infinity in harmonic $m$ comes 
almost entirely from the moments of lowest allowed orders $l$, with the current
moments of order $l$ being, \'a priori, comparable to the mass moments of
order $l+1$. 

For $m=1$, the lowest allowed order for either mass or current is quadrupolar,  
since gravitational waves are always quadrupolar or higher. For circular
orbits in the equatorial plane, the $m=\pm1$ components of the mass 
quadrupole moment vanish, so the dominant waves are current 
quadrupolar $S^{2,\pm1}$ and mass octupolar $I^{3,\pm1}$. 
All other multipolar contributions to $\dot E_{\infty 1}$
are smaller than these by at least $(\Omega r)^2 = \tilde\Omega^{2/3}$. 
The contributions of $S^{2,\pm1}$ to the radiated power $\dot E_{\infty 1}$ can
be derived from Eqs.\ (4.16), (5.27), (2.18a,c), (2.7) and (2.8)
of Ref.~\cite{rmp}; and those of $I^{3,\pm1}$, 
from Eqs.\ (4.16), (5.27), (2.7), and (2.8) of
\cite{rmp}.  The sum of these dominant contributions 
is
\begin{equation}
\dot E_{\infty 1} = {5\over28}\eta^2 \tilde\Omega^4{\cal E}_{\infty 1}\;,
\label{dotE1}
\end{equation} 
where we have tacked on the general relativistic correction factor 
$\dot{\cal E}_{\infty 1}$ to account for contributions from all the
higher-order multipoles and to make the formula be valid not just for large
orbital radii $r$ but for all $r\ge r_{\rm isco}$.  In Eq.\ (\ref{dotE1}), the
numerical factor is $5/28 = 8/45 + 1/1260$, where the big piece $8/45$ 
is current quadrupolar, while the tiny piece $1/1260$ is mass octupolar.

For harmonic $m\ge2$, the lowest allowed multipoles are of order $l=m$,
and the mass moment $I^{m,\pm m}$ is nonzero so it dominates.
All other multipolar contributions to $\dot E_m$ 
are down from these by at least 
$\tilde\Omega^{2/3}$.  An expression for $\dot E_m$
can be derived from Eqs.~(4.16), (5.27), (2.7), and (2.8) of 
Ref.~\cite{rmp}.  The result is
\FL
\begin{eqnarray}
\dot E_m = && {2(m+1)(m+2)(2m+1)!m^{2m+1} \over (m-1)[2^m m!
(2m+1)!!]^2} \nonumber\\
&& \times \eta^2\tilde\Omega^{2+2m/3}{\dot {\cal E}}_{\infty m},
\label{dotEm}
\end{eqnarray}
where $\dot {\cal E}_{\infty m}$ is the relativistic correction and $(2m+1)!! \equiv
(2m+1)(2m-1)(2m-3)\cdots 1$.  
For $m=2$, $3$, and $4$, this expression reduces to
\begin{eqnarray}
\dot E_2 =& {32\over 5} \eta^2\tilde\Omega^{10/3}{\dot {\cal E}}_{\infty 2},
\nonumber\\
\dot E_3 =& {243\over 28} \eta^2\tilde\Omega^{4}{\dot {\cal E}}_{\infty 3}, 
\label{dotE234} \\
\dot E_4 =& {8192\over 567} \eta^2\tilde\Omega^{14/3}{\dot {\cal E}}_{\infty 
4}. \nonumber
\end{eqnarray}
Note that the low-$\tilde\Omega$ limit of $\dot E_2$ is identical to that 
of the total energy loss $\dot E_{\rm GW}$ [Eq.~(\ref{dotE})], as it must be
since the $m=2$ harmonic dominates at low orbital velocities.

From Eqs.~(\ref{dotE1}) and (\ref{dotEm}) for $\dot E_m$ and the general 
relationship
(\ref{ampenergy}) between the waves' amplitude and energy, we obtain
the following Newtonian-order expression for the amplitude in harmonic
$m$:
\begin{mathletters}
\begin{eqnarray}
h_{o,1} = && \sqrt{5\over7} {\eta M\over r_o} \tilde\Omega {\cal H}_{o,1},
\label{ho1}\\
h_{o,m} = && \sqrt{8(m+1)(m+2)(2m+1)!m^{2m-1} \over (m-1)[2^m m!
(2m+1)!!]^2} \nonumber\\
&& \times {\eta M\over r_o} \tilde\Omega^{m/3}{\cal H}_{o,m} \quad\hbox{for
}m\ge2,
\label{hom}
\end{eqnarray}
\label{hom1}
\end{mathletters}
where the relativistic correction is related to that for the energy by 
\begin{equation}
{\cal H}_{o,m} = \sqrt{\dot {\cal E}_{\infty m}}.
\label{Hom}
\end{equation}
For the dominant, $m=2$, radiation Eq.~(\ref{hom}) becomes
\begin{eqnarray}
&&h_{o,2} = \sqrt{32\over5}{\eta M\over r_o}\tilde\Omega^{2/3}{\cal
H}_{o,2} \nonumber\\
&&={3.6\times10^{-22}\over r_o/1{\rm Gpc}} \left({\mu\over 10M_\odot}\right)
\left({M\over 10^6M_\odot}\right)^{2\over3} 
\left({f_2\over .01{\rm Hz}}\right)^{2\over3}{\cal
H}_{o,2} \nonumber\\
&&={3.6\times10^{-23}\over r_o/1{\rm Gpc}} \left({\mu\over M_\odot}\right)
\left({M\over 100M_\odot}\right)^{2\over3} 
\left({f_2\over 100{\rm
Hz}}\right)^{2\over3}{\cal H}_{o,2}\;. \nonumber\\
\label{ho2} 
\end{eqnarray}
From Eqs.~(\ref{hom1}) for $h_{o,m}$, the definition of $h_{c,m}$ in
terms of $h_{o,m}$, the relation $f_m=(m/2\pi)\Omega$, and
Eq.~(\ref{omega2overdotomega}), we obtain the following expression for 
the characteristic amplitude in harmonic $m$
\begin{mathletters}
\begin{eqnarray}
h_{c,1} = && {5\over\sqrt{672\pi}} {\eta^{1/2} M\over r_o}
\tilde\Omega^{1/6}{\cal H}_{c,1},
\label{hc1}\\
h_{c,m} = && \sqrt{5(m+1)(m+2)(2m+1)!m^{2m} \over 12\pi (m-1)[2^m m!
(2m+1)!!]^2} \nonumber\\
&& \times {\eta^{1/2} M\over r_o} \tilde\Omega^{(2m-5)/6}{\cal H}_{c,m} \quad
\hbox{for } m\ge2,
\label{hcm}
\end{eqnarray}
\label{hcm1}
\end{mathletters}
where [using Eq.~(\ref{Hom})] the relativistic correction is related to
earlier ones by
\begin{equation}
{\cal H}_{c,m} = \sqrt{{\cal N}\dot {\cal E}_{\infty m}}.
\label{Hcm}
\end{equation}
For $m=2$, expression (\ref{hcm}) becomes
\begin{eqnarray}
&&h_{c,2} = \sqrt{2\over3\pi} {\eta^{1/2}M\over r_o\tilde\Omega^{1/6}}{\cal
H}_{c,2} \nonumber\\
&&={1.0\times10^{-19}\over r_o/1{\rm Gpc}}
\left({\mu\over10M_\odot} \right)^{1\over2}
\left({M\over10^6M_\odot}\right)^{1\over3}
\left({.01{\rm Hz}\over
f_2}\right)^{1\over6}
{\cal H}_{c,2}\nonumber\\
&&={3.2\times10^{-22}\over r_o/1{\rm Gpc}}
\left({\mu\over M_\odot} \right)^{1\over2}
\left({M\over100M_\odot}\right)^{1\over3}
\left({100{\rm Hz}\over
f_2}\right)^{1\over6}
{\cal H}_{c,2}\;.\nonumber\\
\label{hc2}
\end{eqnarray}

All of the relativistic correction functions can be expressed analytically
in terms of $\dot {\cal E}$, and $\dot {\cal E}_{\infty m}$.  
This has almost been done
already:  The correction functions $\cal T$, ${\cal H}_{o,m}$ and ${\cal
H}_{c,m}$ have been expressed in terms of $\dot {\cal E}$, $\dot {\cal
E}_{\infty m}$, $\cal N$ and $\tilde\Omega_{\rm isco}$  
by Eqs.~(\ref{calT}), (\ref{Hom}) and (\ref{Hcm}) respectively.
All that remains is to derive an expression for $\cal N$ in terms of $\dot 
{\cal E}$, and an expression for $\tilde\Omega_{\rm isco}$. 

The derivations are based on the Kerr-metric relations
\begin{equation}
E = - \eta M {1-2/\tilde r + a/\tilde r^{3/2} \over \sqrt{1-3/\tilde r +
2a/\tilde r^{3/2}}}
\label{Er}
\end{equation}
for the object's total energy in terms of its dimensionless
orbital radius $\tilde r$ (Eq.~(5.4.7b) of Ref.~\cite{leshouches}), and
\begin{equation}
\tilde r = (\tilde\Omega^{-1} -a)^{2/3}.
\label{rOmega}
\end{equation}
for its orbital radius in terms of its orbital angular velocity (Eq
(2.16) of Ref.\ \cite{bpt}).  By
differentiating these equations with respect to time
and combining with each other and with Eqs.~(\ref{omega2overdotomega})
and (\ref{dotE}) for $\Omega^2/\dot\Omega$ and $\dot E_{\rm GW}
= -\dot E$, we obtain
\begin{eqnarray}
{\cal N} = && {1\over \dot {\cal E}}
\left ( 1+{a\over\tilde r^{3/2}}\right )^{5/3}
\left ( 1-{6\over \tilde r} + {8a\over\tilde r^{3/2}} - {3a^2\over
\tilde r^2} \right ) \nonumber \\ 
&& \times \left ( 1- {3\over \tilde r} + {2a\over \tilde
r^{3/2}}\right )^{-3/2}.
\label{Nr}
\end{eqnarray}
When $\tilde r$ is regarded as the function (\ref{rOmega}) of
$\tilde\Omega$, this becomes the desired expression for $\cal N$ in
terms of $\dot {\cal E}$ and $\tilde\Omega$. 

The innermost stable circular orbit (isco)
is at the location $\tilde r_{\rm isco}$ where
the object's total energy $E(\tilde r)$ is a minimum, 
or equivalently where
$\dot\Omega$ is infinite, or equivalently where $\cal N$ vanishes; i.e.,
$\tilde r_{\rm isco}$ is that root of the quartic equation
$\tilde r^2 -6\tilde r + 8a\tilde r^{1/2} -3a^2 =0$
which lies between 1 (when $a=1$) and 6 (when $a=0$).  An analytic
expression for $r_{\rm isco}$ has been given by Bardeen, Press, and
Teukolsky\cite{bpt}:
\begin{eqnarray}
\tilde r_{\rm isco} = &&3+Z_2-{\rm sign}(a)[(3-Z_1)(3+Z_1+2Z_2)]^{1/2}
\;,\nonumber\\
&&Z_1\equiv 1+(1-a^2)^{1/3}[(1+a)^{1/3} + (1-a)^{1/3}]\;,\nonumber\\
&&Z_2\equiv (3a^2 + {Z_1}^2 )^{1/2}\;. \label{rms}
\end{eqnarray}
The dimensionless
orbital angular velocity at the isco
$\tilde\Omega_{\rm isco}$ is expressed in terms
of this $\tilde r_{\rm isco}$ by Eq.\ (\ref{rOmega}):
\begin{equation}
\tilde\Omega_{\rm isco} = {1\over {\tilde r_{\rm isco}}^{3/2} +a}\;.
\label{Omegams}
\end{equation} 

We note, in passing, approximate analytic formulae for the relativistic
corrections $\cal T$ and ${\cal N}_{\rm orb}$ to the time $T$ and number of
remaining orbits ${\cal N}_{\rm orb}$ until the end of inspiral---for the
special case of a nonspinning black hole, $a=0$.  Inserting expression
(\ref{Nr}) into Eqs.\ (\ref{calT}) and (\ref{calNorb}), and noting
from Table \ref{ScrDotE.tbl} that in each of these equations
$\dot{\cal E}$ is a much more slowly varying
function of the integration variable than the rest of the integrand, 
we pull $\dot{\cal E}$ out of the integral (i.e., we perform the first
step of an integration by parts) and then perform the integration analytically.
The results are
\begin{eqnarray}
{\cal T} &\simeq& 
{1\over\dot{\cal E}} \left[ {1-{7\over2}u
- {147\over8}u^2 - {2205\over 16}u^3 + {19845\over16}u^4
\over\sqrt{1-3u}}\right.
\nonumber\\
&&\left. -{4671\over 8\sqrt{2}}u^4 +
{19845\over32}u^4\ln\left({3(1+\sqrt2)^2u\over(1+\sqrt{1-3u})^2}
\right)\right]\;,
\label{calTanalytical}
\end{eqnarray}
\begin{equation}
{\cal N}_{\rm orb} \simeq {1\over\dot{\cal E}}\left(
{1-4u-48u^2+288u^3\over\sqrt{1-3u}} -24\sqrt{3u^5}\right)\;,
\end{equation}
where $u=\tilde\Omega^{2/3} = 1/\tilde r$.  These formulae agree
with the numerical values of $\cal T$ and ${\cal N}_{\rm orb}$
in Tables \ref{ScrT.tbl} and \ref{ScrNorb.tbl} below to within 3 per cent
at $6.02 < \tilde r < 18$, and to within 1 per cent at $6\le\tilde r<6.02$ and
$\tilde r > 18$.

\section{Tables of Relativistic Correction Functions}
\label{sec:Tables}

We shall use two dimensionless parameters to measure the distance of an orbit
from the isco: the ratio $r/r_{\rm isco}\equiv \tilde r / \tilde r_{\rm isco}$ 
of the orbit's Boyer-Lindquist
radial coordinate $r$ to its value at the isco, and the ratio
$\Omega/\Omega_{\rm isco} \equiv \tilde\Omega/\tilde\Omega_{\rm isco}$ of the 
orbit's angular velocity to that at the isco.  The relationship between these
two parameters is given by Eqs.\ (\ref{rOmega}), (\ref{rms}) and
(\ref{Omegams}), and is tabulated in Table \ref{Omega.tbl}.

We have integrated the Teukolsky-Sasaki-Nakamura 
equation for
perturbations of a Kerr black hole to obtain the functions $\dot {\cal E}
(\tilde\Omega)$ and $\dot {\cal E}_{\infty m} (\tilde\Omega)$, 
and we have then used 
Eqs.~(\ref{rOmega}), (\ref{Nr}), (\ref{calT}) and 
(\ref{calNorb}) to compute 
$\cal N$, $\cal T$ and ${\cal N}_{\rm orb}$.  These 
functions are listed in Tables \ref{ScrDotE.tbl}--\ref{ScrNorb.tbl}
and some of our numerical methods are described in the Appendix.
As a byproduct of these calculations, we have inferred what fraction $\dot
E/\dot E_{\rm GW}$ of the total
rate of energy emission goes down the hole's horizon; that fraction is shown in
Table \ref{DotEHOverDotEtot.tbl}. 

\section{Applications to LISA}
\label{sec:Applications}

\subsection{LISA Noise}

Tentative error budgets for LISA are spelled out in Tables 4.1 and 4.2 of
the LISA Pre-Phase-A Report \cite{PPA-2}.  Various researchers have computed
LISA noise spectra from those error budgets.  It is conventional, for LISA,
to characterize the noise by the sensitivity to periodic sources for one year
integration time and a signal-to-noise ratio of 5, averaged over source
directions and polarizations (``sky averaged''). 
We shall denote this quantity by 
$h_{\rm SN5,1yr}^{\rm SA}$.  It is related to the sky-averaged spectral
density introduced in the paragraph before Eq.\ (\ref{ampenergy}) by
$h_{\rm SN5,1yr}^{\rm SA} = 5 \sqrt{S_h^{\rm SA}\Delta f}$, where $\Delta f =
1/1{\rm yr}$ is the bandwidth for the one-year integration time; and
correspondingly, it is related to the sky-averaged rms noise in a 
bandwidth equal to frequency, $h_n(f) = \sqrt{fS_h^{\rm SA}(f)}$ (which we use
in this paper), by
\begin{equation} 
h_n(f) = {1\over 5}{\sqrt{f\over\Delta f}h_{\rm SN5,1yr}^{\rm SA}}(f)\;.
\label{hnh1yr}
\end{equation}
We have deduced $h_n(f)$ using this equation and the values of $h_{\rm
SN5,1yr}^{\rm SA}(f)$ computed by various 
researchers\cite{mdt,aet,schilling}; we
plot it as a thick solid curve in Figs.\  
\ref{fig:hc1million}--\ref{fig:hc10tenmillion} below.

It is likely that LISA's performance will be compromised at 
$f\alt 0.003$ Hz by a stochastic background due to white-dwarf
binaries.  The most recent estimate of that stochastic background
is by Hils and Bender \cite{hilsbender}; it agrees satisfactorily
with an estimate by Webbink and Han\cite{webbinkhan}.  We have used
a simple piece-wise straight-line fit to the logarithm of the Hils-Bender
white-dwarf-background noise curve: straight lines that join the following
points in $(\log_{10}f, \log_{10}h_{\rm SN5,1yr}^{\rm SA})$ where $f$ is
measured in Hz:
\begin{eqnarray}
&&(-4, -20.518)\;,\quad (-3.62, -20.737)\;,\quad (-2.78, -21.66)\;,
\nonumber\\
&&(-2.61, -22.90)\;, \quad (-2, -23.731)\;. 
\label{BHWD}
\end{eqnarray}
This white-dwarf noise, converted to our conventions via Eq.\ 
(\ref{hnh1yr}), is shown as a thick dashed curve in Figs.\ 
\ref{fig:hc1million}--\ref{fig:hc10tenmillion} below. 

\subsection{Detectable Systems}
\label{sec:DetectableSystems}

Of greatest interest, for probing the spacetime geometries of massive black
holes, is the gravitational radiation emitted during the last year of inspiral
of a compact object.  In planning the LISA mission, it is important to know the
detectability of these final-year waves, as a function of the system's
parameters: the hole's mass $M$ and spin $a$, the object's mass $\mu$, and the
distance $r_o$ from Earth.  Previous studies of this issue \cite{mdt,PPA-2}
have assumed that the massive hole is nonspinning, $a=0$. 

It is straightforward to compute the rms signal to noise ratio 
$(S/N)_{\rm rms}$ 
(averaged over detector and system orientations) from
Eq.\ (\ref{SoverN}), using the noise amplitudes $h_n$ described above
and the dominant $m=2$ characteristic amplitude $h_{c,2}$ of 
Eqs.\ (\ref{hc2}) and (\ref{Hcm}), with $\cal N$ and $\dot {\cal
E}_{\infty m}$ taken from Tables \ref{ScrN.tbl} and \ref{ScrDotE2.tbl}.
In this calculation, the frequency $f_2 = \Omega /\pi$ ranges from its
value at time $T = 1$ year [Eq.\ (\ref{T}) and Table \ref{ScrT.tbl}] to
its value at the isco.  

\begin{figure}
\epsfxsize=3.2in\epsfbox{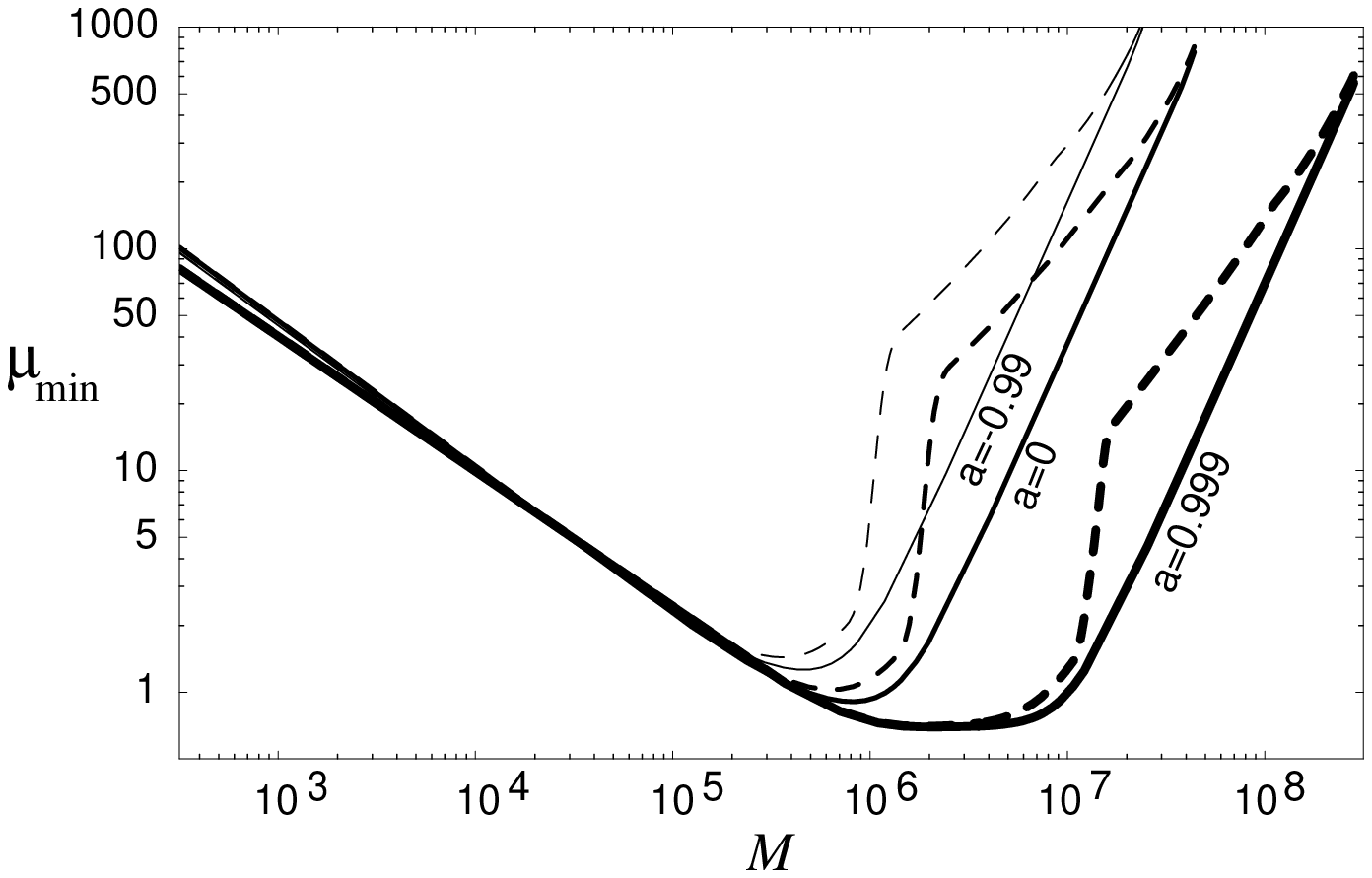}
\caption{ The minimum mass $\mu_{\rm min}$ that the inspiraling object
must have in order to
produce a signal to noise ratio $S/N>10$ in its dominant harmonic,
$m=2$, during the last year of its inspiral. This $\mu_{\rm
min}$ is plotted as a function of the black-hole mass $M$, for various
black-hole spin parameters $a$.  The solid curves are for the LISA noise
spectrum; the dashed curves are for the LISA noise plus a stochastic
background noise due to white-dwarf binaries.\protect\cite{LisaFit}
}
\label{fig:muminyr}
\end{figure}

In view of the complexity of the data analysis for these waves, a
signal to noise ratio of about 10 may be required for their detection, 
and in view of the estimated event rates (Sec.\ \ref{sec:Introduction}),
it is necessary that LISA see out to at least $r_o = 1$Gpc.
Accordingly, we have computed the range of masses $\mu$ and $M$ and
black-hole spins $a$ for  
which $(S/N)_{\rm rms} >10$ at a distance $r_o = 1$ Gpc. This range of 
``detectable systems'' is shown
in Fig.\ \ref{fig:muminyr} for LISA without the white-dwarf background
(solid curves) and with the background (dashed curves).\cite{LisaFit} 

Several features of this figure deserve comment:  
\begin{itemize}
\item Inspiraling white dwarfs and neutron stars ($\mu \alt 1.4
M_\odot$) are barely detectable, with $(S/N)_{\rm
rms} = 10$, at 1Gpc.  It would be highly desirable to reduce LISA's 
design noise floor by a factor two or three, to give greater confidence
of detection.
\item For $\mu = 10M_\odot$ inspiraling black holes,
the detectable systems have a wide range of central black-hole
masses: $10^4 M_\odot \alt M \alt 10^7 M_\odot$.
\item The upper limit on detectable central-hole masses $M$ depends strongly on
the black-hole spin: for $\mu = 10M_\odot$ it ranges from $2\times10^6M_\odot$
to $3\times 10^7 M_\odot$ without the white-dwarf background, and
$1\times 10^6$ to $1.5\times 10^7$ with the background.  (The spins shown
are for no rotation $a=0$, and for near the maximum rotation,
$a\simeq\pm 0.998$, that can be produced by spinup via accretion from a disk
\cite{max_spin}.) 
\item The white dwarf background reduces the maximum detectable black
hole mass by about a factor 2.5, independent of the spin.
\item The white dwarf background and the black hole spin have little
influence on the minimum detectable mass $M$.  This is because, at low
$M$ the object travels a large radial distance in its last year of
life, so most of the signal to noise comes from radii $r \gg r_{\rm isco}$
where the spin is unimportant, and (by virtue of the small $M$) most
comes from frequencies high enough that the white-dwarf background is 
negligible.
\end{itemize}

\begin{figure}
\epsfxsize=3.2in\epsfbox{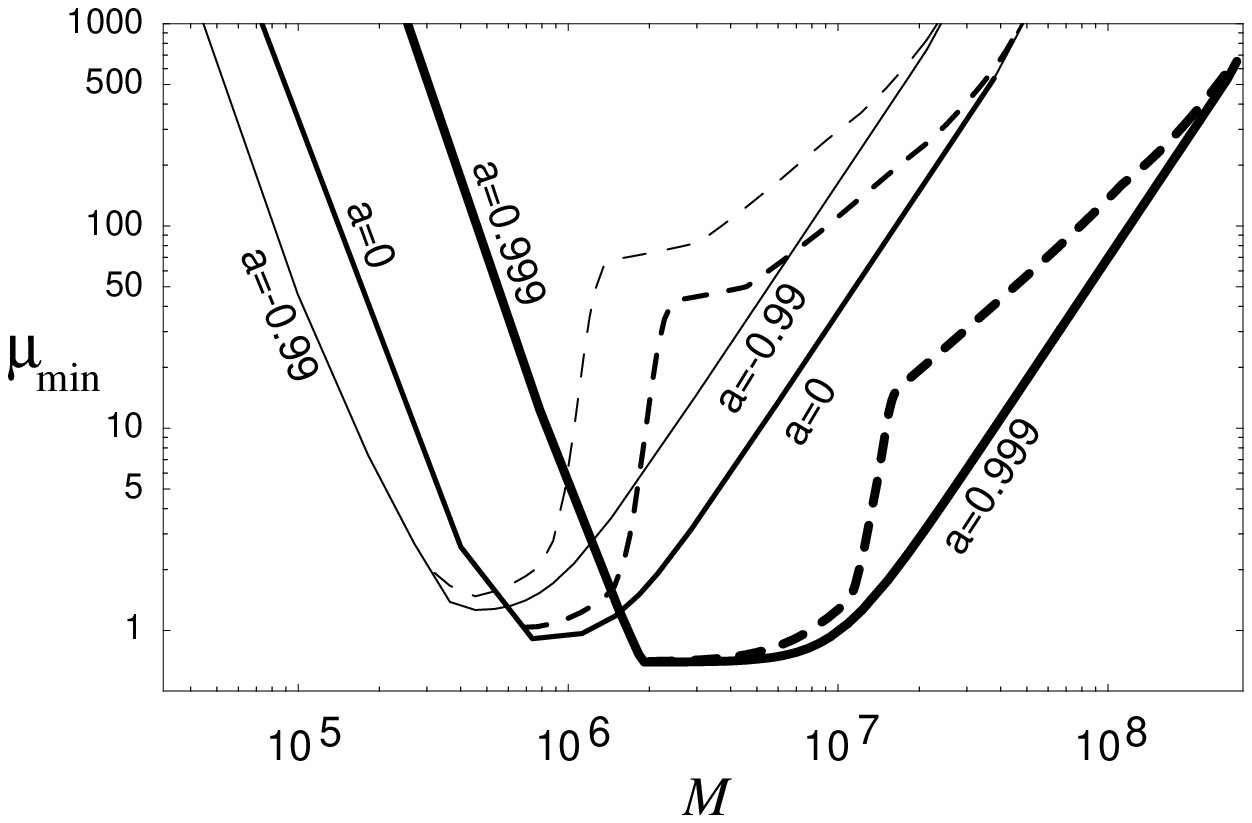}
\caption{ The minimum mass $\mu_{\rm min}$ that the inspiraling object 
must have in order to 
produce a signal to noise ratio $S/N>10$ in its dominant harmonic,
$m=2$, during the last year of its inspiral, {\it and in the vicinity of the
horizon,} $(2/3)f_{2,{\rm isco}} < f_2 < f_{2,{\rm isco}}$.  This $\mu_{\rm
min}$ is plotted as a function of the black-hole mass $M$, for various
black-hole spin parameters $a$.  The solid curves are for the LISA noise
spectrum; the dashed curves are for the LISA noise plus a stochastic
background noise due to white-dwarf binaries.\protect\cite{LisaFit} 
}
\label{fig:muminisco}
\end{figure}

For probing the immediate vicinity of the horizon, we are interested in
waves with frequencies, say, $(2/3)f_{2,{\rm isco}} < f_2 < f_{2,{\rm
isco}}$.  Figure \ref{fig:muminisco} shows the range of 
systems for which $(S/N)_{\rm rms} > 10$ in this frequency band, at
a distance $r_o = 1$Gpc, during the last year of inspiral.  Note that
restricting attention to this near-horizon frequency
range has reduced substantially the set of detectable systems:  for $\mu =
10 M_\odot$, the minimum
black-hole mass is increased by a factor $\sim 20$ to $100$, depending
on the spin $a$.  Nevertheless, there is still a wide range of systems
accessible for study.

For distances larger than $r_o \sim 1$ Gpc, cosmological effects have a
significant influence on the signal \cite{finn_cosmology}.  At fixed
$\{a,M(1+z),\mu(1+z)\}$ (where $z$ is the cosmological redshift), the
characteristic amplitude and signal to noise ratio scale $\propto
1/r_{oL}$, where $r_{oL}$ is the luminosity distance to Earth.  The
scaling of $(S/N)_{\rm rms}$ with $\mu(1+z)$ is not so
simple, because it influences the waves' frequency 
evolution in complicated ways that entail the relativistic correction
functions.  For extremely rough estimates, one can use the leading-order
(in $\tilde\Omega$) expression for $h_{c,2}$ [Eq.\ (\ref{hc2}), $S/N
\propto h_{c,2} \propto \mu^{1/2} / r_{\rm oL}$] to infer $\mu_{\rm min}
(1+z) \propto r_{oL}^2$ for the minimum detectable object mass at fixed
$a$ and $M(1+z)$,  
but for reliable results,
one must repeat the analysis (sketched above) by which we arrived at
Figs.\ \ref{fig:muminyr} and \ref{fig:muminisco}.

\subsection{Evolution of the Waves During Inspiral}
\label{sec:Evolution}

\begin{figure}
\epsfxsize=3.3in\epsfbox{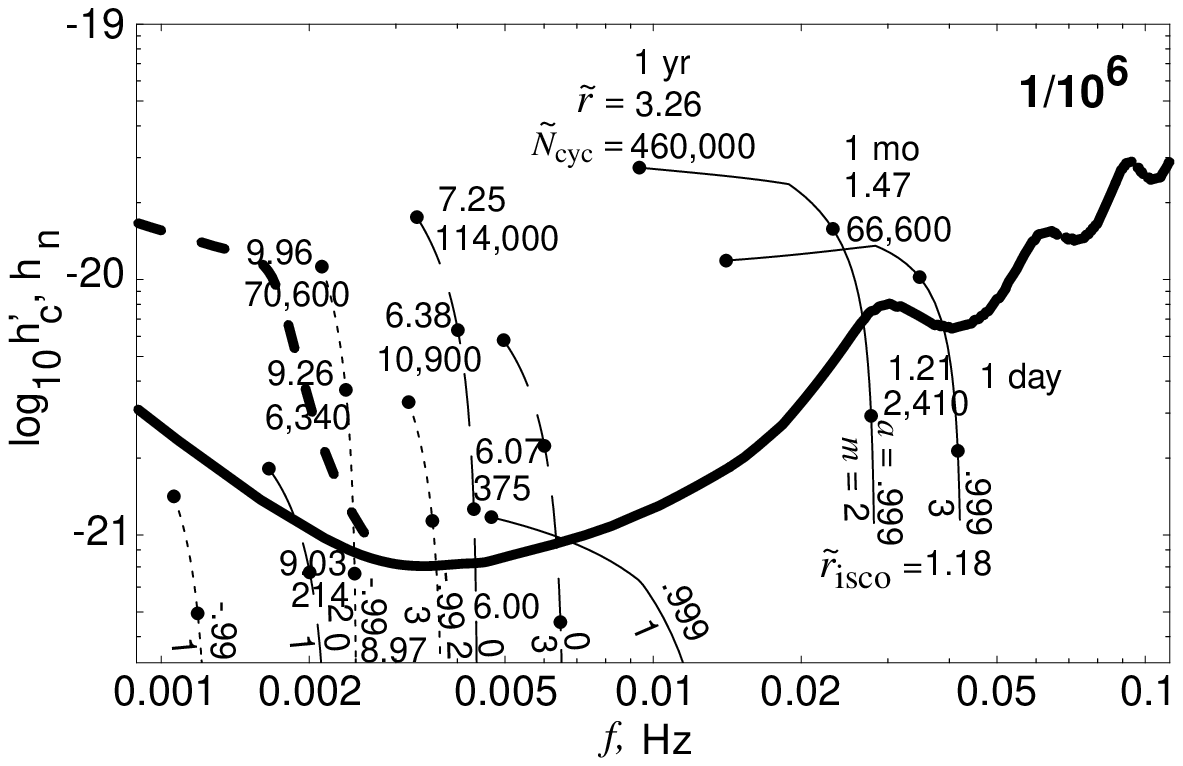}
\caption{Gravitational waves from a $1 M_\odot$ white dwarf or neutron star
spiraling into
a $10^6 M_\odot$ black hole at 1Gpc distance from Earth, as observed by LISA.  
The thick solid curve is LISA's rms noise level $h_n(f)$
averaged over the sky; the thick dashed curve is an estimate of the 
stochastic-background ``noise'' produced by white dwarf binaries. 
Each thin curve is the modified characteristic amplitude
$h_c'(f)$ for a harmonic of the waves, and is labeled 
vertically by the hole's spin parameter $a$ and the harmonic number $m$.  
The three dots on each curve indicate the waves properties
one year (left dot), one month (center dot) and one day (right dot) 
before reaching
the isco.  The dots on the dominant, $m=2$, harmonics are labeled by the
orbital radius $\tilde r = r/M$ and the number of $m=2$
wave cycles remaining until 
the isco.  The isco radius is shown at the bottom of each $m=2$ curve.
}
\label{fig:hc1million}
\end{figure}

\begin{figure}
\epsfxsize=3.3in\epsfbox{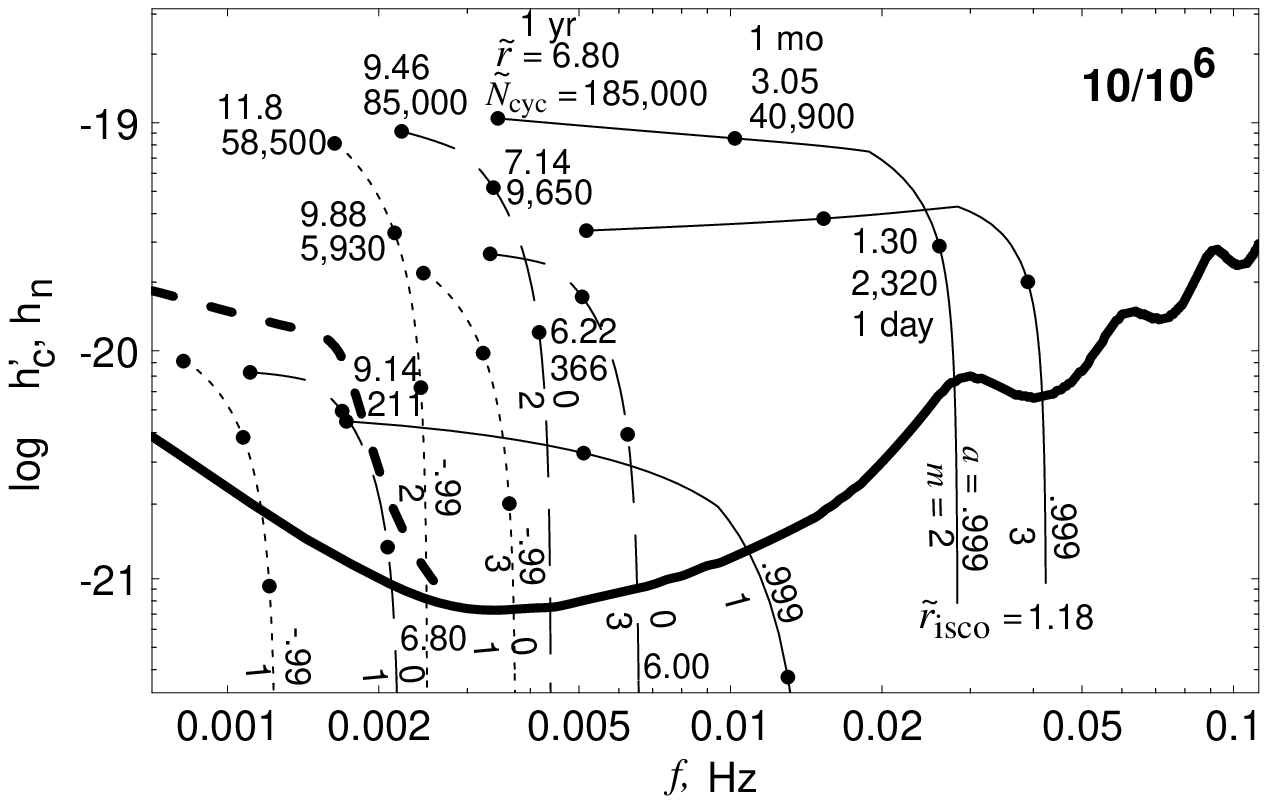}
\caption{Gravitational waves from a $10M_\odot$ black hole spiraling into 
a $10^6 M_\odot$ black hole at 1Gpc distance from Earth, as observed by
LISA. For notation see caption of Fig.\ \protect\ref{fig:hc1million}. 
}
\label{fig:hc10million}
\end{figure}

\begin{figure}
\epsfxsize=3.3in\epsfbox{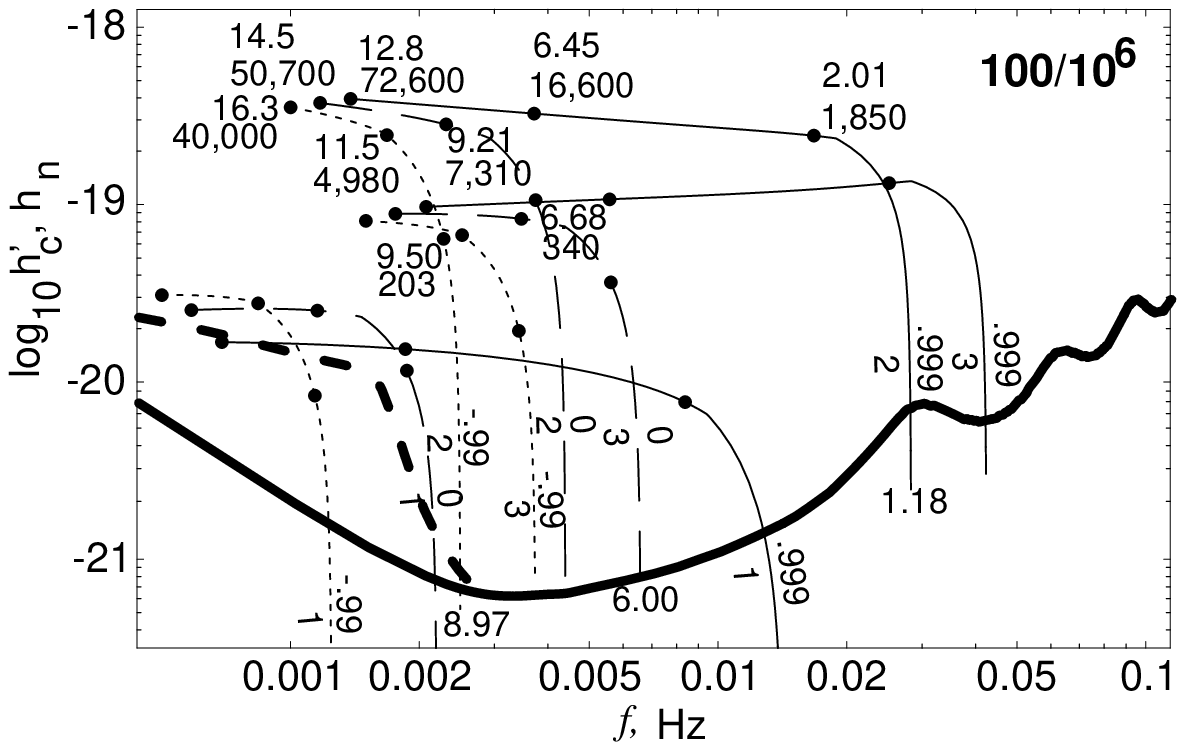}
\caption{Gravitational waves from a $100 M_\odot$ black hole
spiraling into
a $10^6 M_\odot$ black hole at 1Gpc distance from Earth, as observed by
LISA.  For notation see caption of Fig.\ \protect\ref{fig:hc1million}.
}
\label{fig:hc100million}
\end{figure}

\begin{figure}
\epsfxsize=3.3in\epsfbox{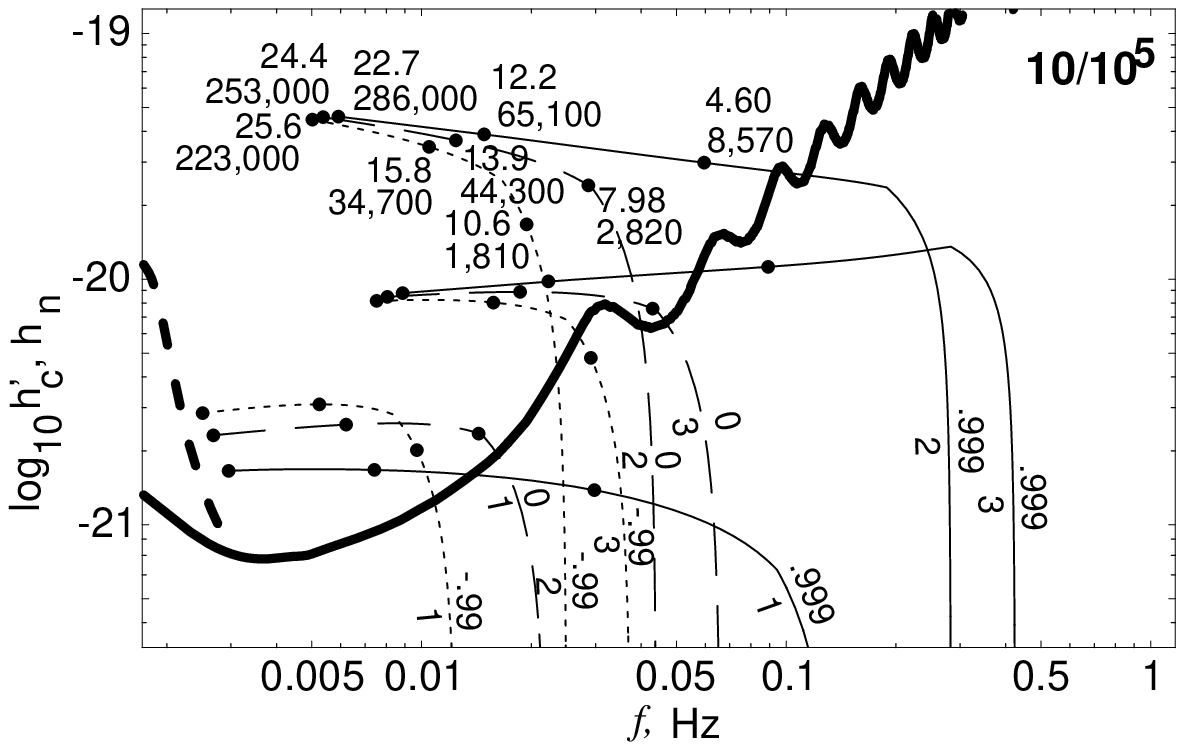}
\caption{Gravitational waves from a $10 M_\odot$ black hole
spiraling into
a $10^5 M_\odot$ black hole at 1Gpc distance from Earth, as observed by
LISA.  For notation see caption of Fig.\ \protect\ref{fig:hc1million}.
}
\label{fig:hc10hundredthousand}
\end{figure}

\begin{figure}
\epsfxsize=3.3in\epsfbox{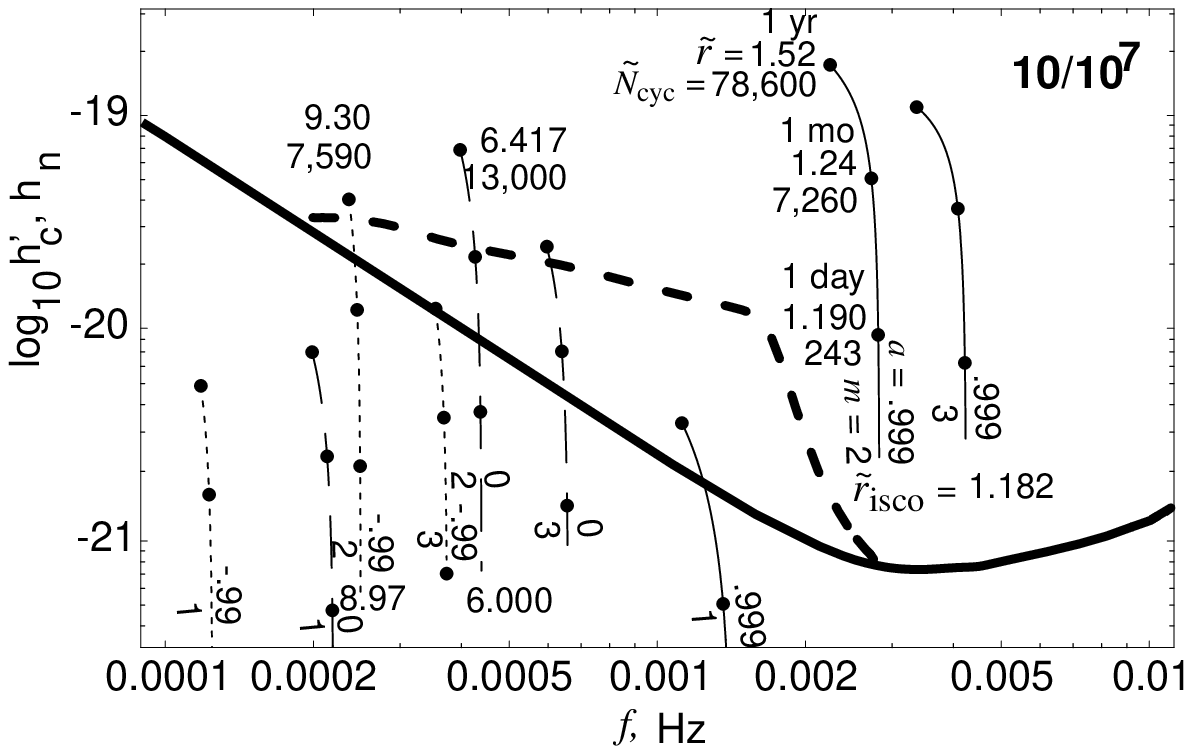}
\caption{Gravitational waves from a $10 M_\odot$ black hole
spiraling into
a $10^7 M_\odot$ black hole at 1Gpc distance from Earth, as observed by
LISA.  For notation see caption of Fig.\ \protect\ref{fig:hc1million}.
}
\label{fig:hc10tenmillion}
\end{figure}

To gain insight into the emitted waves and how they evolve during the
inspiral, we have constructed Figs.\ 
\ref{fig:hc1million}---\ref{fig:hc10tenmillion}.  Each figure depicts
the waves' evolution for the value of object mass $\mu$ and hole mass $M$
(in solar masses) listed in bold letters in the upper right corner.
The horizontal axis is the waves' frequency $f$ and the vertical axis,
their modified characteristic amplitude $h'_c$.  As the inspiral proceeds,
the waves sweep upward in frequency (left to right) along one of the thin
curves.  These evolutionary curves are shown for three different values
of the black-hole spin, $a= -0.99$ (retrograde orbit; short-dashed curves), 
$a=0$ (no rotation; long-dashed curves) and $a=+0.999$ (prograde orbit;
solid curves).  For each spin, three curves are
shown corresponding to the three lowest harmonics $m=1,2,3$ of the
orbital frequency.  The values of $a$ and $m$ for each evolutionary
curve are listed near the vertical endpoint of the curve.
Also shown in each figure is the rms noise amplitude 
$h_n$ for LISA: a thick solid curve in the absence of a
white-dwarf-binary background and a thick dashed curve including that
background.

The range of frequency sweep is strongly dependent on the masses $\mu$
and $M$ of object and hole.  Neglecting the relativistic correction
factor $\cal T$ (which is unimportant for this purpose when the
frequency sweep is substantial), Eq.\ (\ref{T}) tells us that $ f_{\rm isco} 
/ f_{1 {\rm yr}}\propto (\mu/M)^{3/8}(1/ M)^{3/8}$, where $f_{1 {\rm yr}}$
is the frequency one year before reaching the isco.  Thus, the
greatest frequency sweep is for the least extreme mass ratio and the smallest
hole mass, $\mu/M = 10/10^5$ (Fig.\ \ref{fig:hc10hundredthousand}) with
$f_2$ sweeping from $\sim 0.006$ Hz to $0.4$ Hz; while
the smallest sweep is for the most extreme mass ratio and largest hole mass,
$\mu/M = 10/10^7$ (Fig.\ \ref{fig:hc10tenmillion}) with $f_2$ sweeping 
only from $\sim 0.0023$ to $0.0027$ Hz.

The height of a signal curve $h'_c$ above the noise curve $h_n$ is about
equal to the signal to noise ratio in an appropriate bandwidth $\Delta
f$: $\Delta f = f$ well away from the endpoint of inspiral, and 
$\Delta f = 2(f-f_{\rm isco})$ near the endpoint; cf.\
the discussion of the definition of $h'_c$ at the end of Sec.\
\ref{sec:Notation}.
Near the endpoint of inspiral $h'_c$ plunges for three reasons: 
(i) because of the narrowing of our chosen bandwidth; 
(ii) because the rate of frequency sweep speeds up due to
flattening of the effective potential for the object's radial motion, 
and this produces
a reduction in the number of cycles $N_{\rm cyc}$ in a given bandwidth
and reduction in $h'_c \propto \sqrt{N_{\rm cyc}}$; (iii) because,
for large $a$
and prograde orbits, the orbit sinks deep into the throat of the hole's
embedding diagram, from where waves have difficulty escaping.

On each signal curve there are three solid dots.  They label
$(f,h'_c)$ for specific times during the inspiral: $T =1$ year before
the endpoint (leftmost dot), $T=1$ month before the end (center dot), and
$T=1$ day before the end (right dot).  Beside the dots for the dominant
harmonic, $m=2$, are shown two numbers that characterize the orbit and
waves at that time: the radius $\tilde r = r/M$
of the orbit in units of the black-hole mass, and the
number of gravitational-wave cycles in the $m=2$ harmonic, from that
time until the endpoint of inspiral.  At the bottom end of each $m=2$ curve is
shown the radius $\tilde r$ of the isco.

It is worthwhile to scrutinize the details of these figures, including
the numbers beside the dots.   Consider, for example, Fig.\
\ref{fig:hc10million} for a $\mu=10M_\odot$ object (black hole) 
spiraling into 
a $M=10^6 M_\odot$ hole.  If the big hole is rapidly rotating and the
orbit is prograde so $a=+0.999$, then the dominant $m=2$ evolutionary
curve shows the object, one year before its death, at $\tilde r = 6.80$
(3.4 Schwarzschild radii), with a signal to noise ratio of $h_c/h_n
\sim 100$, and with 185,000 cycles of gravitational waves left until
death.  One month before death, the object is at $\tilde r =3.05$ (1.53
Schwarzschild radii), with $h_c/h_n \sim 50$, and with 40,000 cycles
left.  One day before death, it is at $\tilde r = 1.30$ (compared to
1.18 for the isco), with $h_c/h_n \sim 10$ and with 2,320 cycles left. 
It is impressive how long the object lingers in the vicinity of the
horizon, and how many wave cycles it emits.

For a nonspinning hole $a=0$, the numbers are less impressive but still
remarkable: the last year is spent spiraling from $\tilde r = 9.46$
(4.73 Schwarzschild radii) to the isco at $\tilde r = 6$ (3
Schwarzschild radii), during which 85,000 wave cycles are emitted and
$h_c/h_n$ drops from $\sim 100$ to $\sim 10$ at one day and then to zero. 

The large number of wave cycles carry a large amount of information
about the source.  We shall discuss this issue in Sec.\ 
\ref{subsec:Information} below.

Figures \ref{fig:hc1million}--\ref{fig:hc100million} illustrate the
influence of the mass of the inspiraling object on the signal strength.  
For $M$ fixed at $10^6 M_\odot$, one year before merger the 
$m=2$ signal to noise ratios $h_c/h_n$ 
are $\sim 15$ for $\mu = 1 M_\odot$, $\sim
100$ for $\mu = 10 M_\odot$ and $\sim 500$ for $\mu = 100 M_\odot$.
This is a moderately faster growth than our crude estimate $\propto
\mu^{1/2}$ in Sec.\ \ref{sec:DetectableSystems}.  Notice that 
$h_c/h_n$ 
drops below 10 one month before the endpoint for $\mu
= 1M_\odot$, and one day before the end for $\mu = 10M_\odot$.

To maximize the exploration of the horizon's vicinity, we want the
object to spend its entire last year at radii $\tilde r \alt 10$.  If
the object is a $10M_\odot$ hole, this is the case when $M\agt 10^6
M_\odot$; cf.\ Figs. \ref{fig:hc10million}, \ref{fig:hc10hundredthousand}
and \ref{fig:hc10tenmillion}.  For $M < 10^6 M_\odot$, such exploration
is debilitated by the large frequency sweep; cf.\ Fig.\
\ref{fig:hc10hundredthousand}.  We have previously met this issue in
Sec.\ \ref{sec:DetectableSystems}.

Figure \ref{fig:hc10tenmillion} shows that the white-dwarf-binary
background is a serious issue for hole masses $M\sim 10^7 M_\odot$,
while Figs.\ \ref{fig:hc1million}--\ref{fig:hc10hundredthousand} show
that it is relatively unimportant for $M\alt 10^6M_\odot$.  We have
previously met this in Sec.\ \ref{sec:DetectableSystems}.

\subsection{Information Carried by the Waves}
\label{subsec:Information}

As is well known \cite{last3minutes}, 
the waves' highest accuracy information is carried by the time
evolution of their phase.  For circular, equatorial orbits, where there
is no orbital precession, 
the phase evolution will be the same for all
the harmonics as for the orbit itself, and that phase evolution is
embodied in $d\Phi / d\ln\Omega = \Omega^2/\dot\Omega$.  
Equation (\ref{omega2overdotomega}) shows this quantity, at fixed frequency, to
be proportional to ${\cal N}/M_{\rm chirp}^2$, where
\begin{equation}
M_{\rm chirp} = \mu^{1/2} M^{1/3}
\label{chirp}
\end{equation}
is the system's chirp mass.  Since a year of observations will
typically entail $N_{\rm cycle} \sim 10^5$ cycles of waves, and by the method of
matched filters one can detect a secular shift of one waveform with
respect to another by a small fraction of a cycle \cite{last3minutes},
the ``raw'' precision for measuring the evolution of ${\cal N}/M_{\rm
chirp}$ will be of order $10^{-6}$.

If most of the last year is spent near the horizon, say at frequencies 
$f/f_{\rm isco} = \Omega/\Omega_{\rm isco} \agt 0.1$ (as will usually 
be the case), then
this phase evolution will depend strongly not only on the chirp mass,
but also---through the function ${\cal N}(f/f_{\rm isco})$---on 
the black-hole spin parameter $a$.  

\begin{figure}
\epsfxsize=3.2in\epsfbox{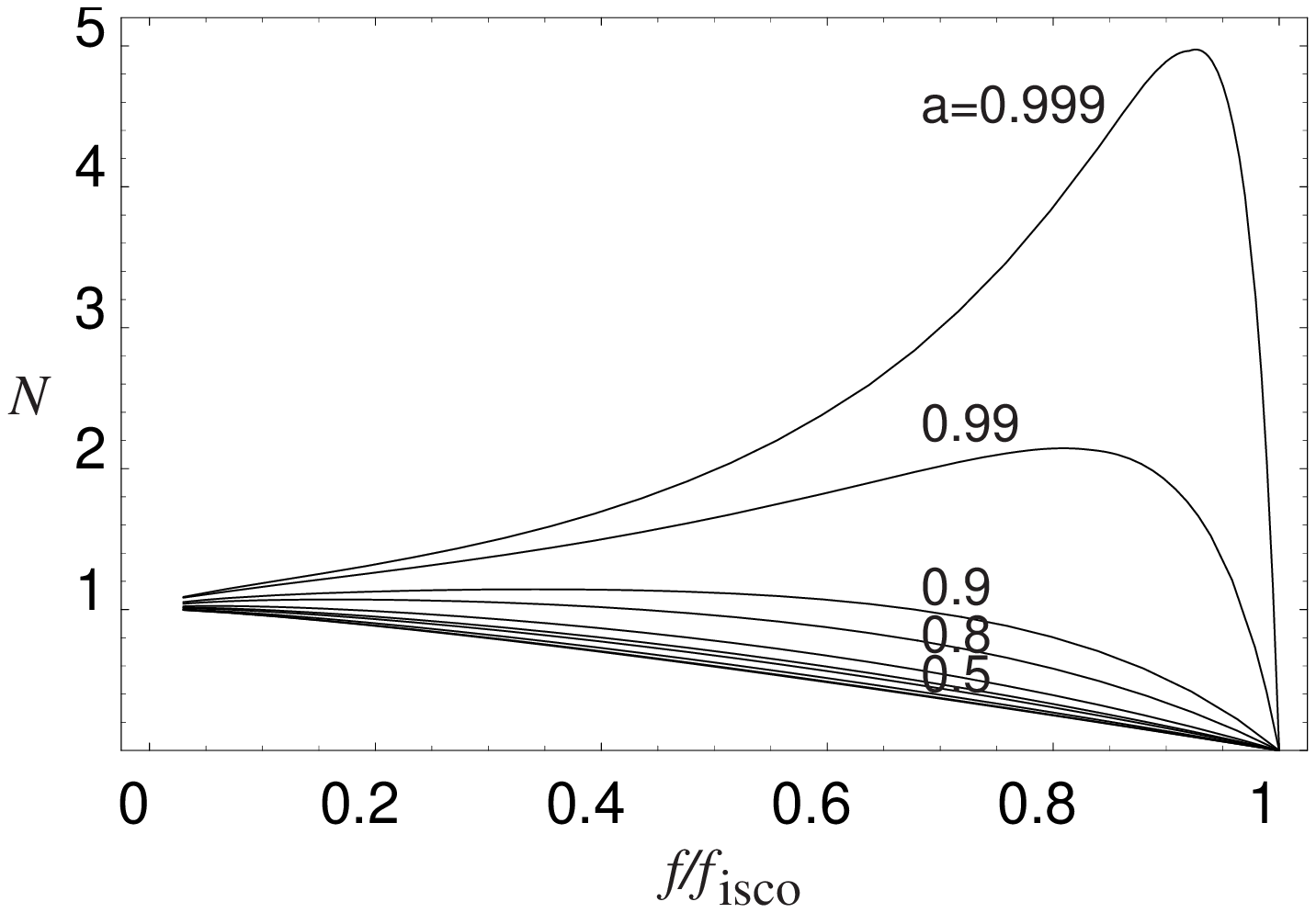}
\caption{ ${\cal N}$ the relativistic correction 
to $d\Phi/d\ln\Omega = \Omega^2/\dot\Omega$ (the number of radians of
orbital inspiral per unit logarithmic change of orbital or gravitational-wave
frequency), plotted against $f/f_{\rm isco} = \Omega/\Omega_{\rm isco}$ (the
ratio of gravitational-wave frequency to the frequency when the isco is reached
and the inspiral ends).  
}
\label{fig:scrN}
\end{figure}

This strong $a$-dependence is exhibited in Fig.\ \ref{fig:scrN}.
Even for $a < 0.5$, where the curves ${\cal N}(f/f_{\rm isco})$ for
different $a$ look very close together, $\partial{\cal N}/\partial a \sim
0.1$, this $a$-dependence translates into $\partial N_{\rm
cycles}/\partial a \sim 10^4$, which is huge.  Thus, it is reasonable
to expect the measured phasing to determine both $a$ and $M_{\rm
chirp}$ to high precision---though a detailed parameter study is needed to be
absolutely certain.

The absolute frequencies associated with the observed phase evolution
(e.g., the measured frequency at the end of inspiral) are determined by
a combination of $a$ and the hole's mass $M$.  This absolute frequency
scale presumably will be measured much less accurately than the phasing
itself, but still, probably, accurately enough to determine the mass
$M$ to a very interesting precision.  Knowing $M_{\rm chirp}$, $a$, and
$M$, one can then compute the object's mass $\mu$; and from the
absolute amplitudes of the waves one can then infer the distance $r_o$
from the system to Earth.

Poisson \cite{poisson} has estimated the accuracies with which such
phase-evolution measurements can determine $M$, $\eta = \mu/M$, and $a$.  His
estimates are based on an analytic model of the signal in which
(translated into our notation) ${\cal N}$ is expanded in powers of
$\Omega - \Omega_{\rm isco}$ and only the leading order term is
kept.  Poisson assumes $M=10^6 M_\odot$, $\mu = 10 M_\odot$, $a \sim 0$
(i.e., not close to $\pm 1$),  and a 
measurement time of one year. For these parameters, our 
Figs.\ \ref{fig:scrN} and \ref{fig:hc10million}
suggest that, for $a\alt0.5$, his expansion may be accurate to within
a few tens of per cent, while for $a\agt 0.9$ it is seriously wrong.
His estimated
measurement accuracies are $\Delta a \sim 0.05/\rho$, $\Delta M/M \sim
0.002/\rho$, and $\delta\eta/\eta \sim 0.06/\rho$, where $\rho$ is the
amplitude signal to noise ratio.

Our Tables \ref{ScrN.tbl} and \ref{ScrT.tbl} for ${\cal N}$ and ${\cal
T}$ (the relativistic corrections to the orbital phase evolution rate
$d\Phi/d\ln f$ and the time $T$ to the end of inspiral) can serve as the
foundation a more definitive computation of the phasing-based measurement
accuracies. 

\begin{figure}
\epsfxsize=3.3in\epsfbox{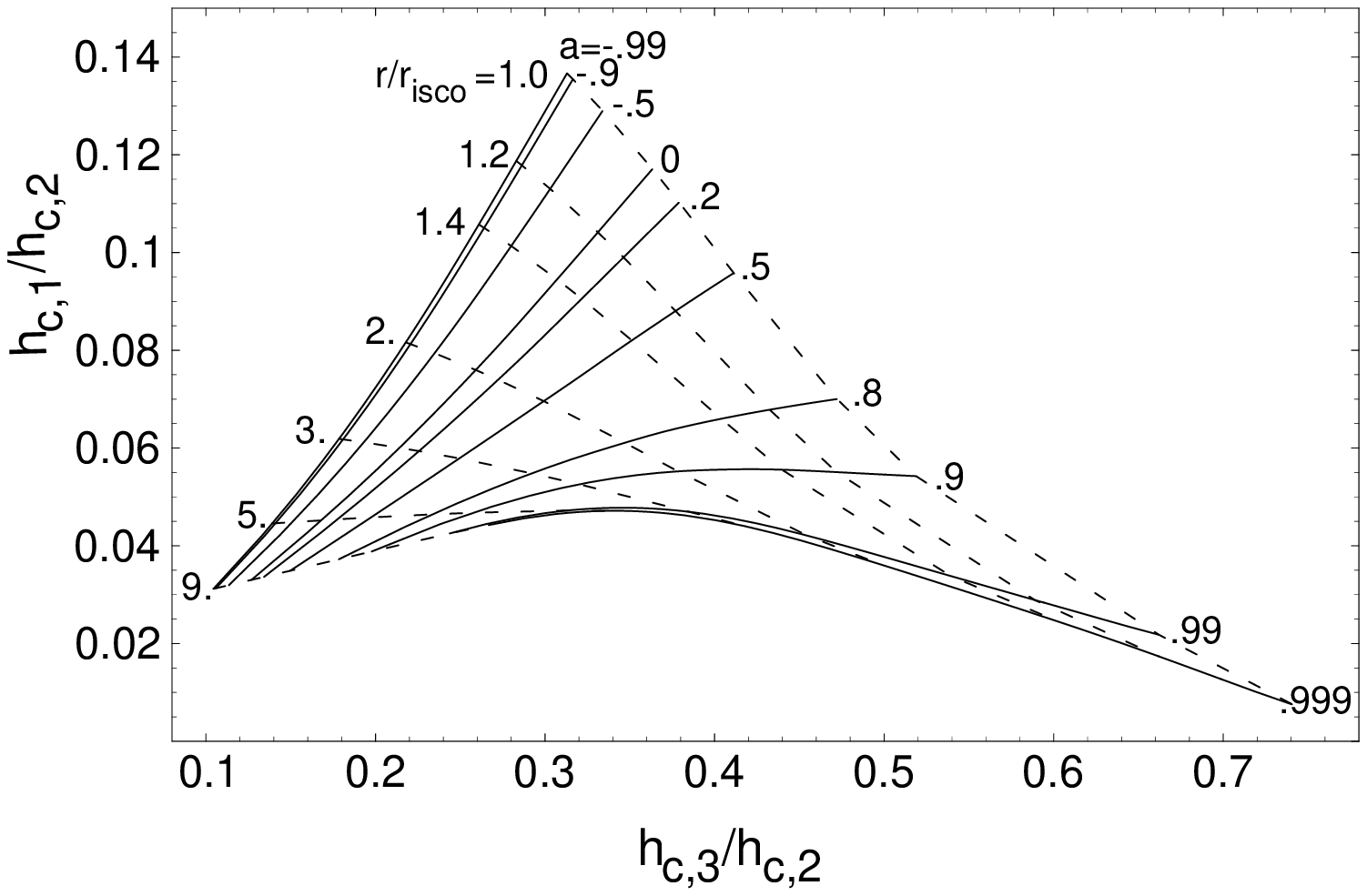}
\caption{The ratios $h_{c,1}/h_{c,2}$ and $h_{c,3}/h_{c,2}$ as functions
of the black hole spin $a$ and the orbiting object's radius $r/r_{\rm isco}$.
}
\label{fig:hc1Overhc2Vshc3Overhc2}
\end{figure}

Information is also carried by the relative amplitudes of the waves'
harmonics.  Most promising, we think, are the amplitude ratios for the
first and second harmonics and for the third and second.  We plot these
ratios in Fig.\  \ref{fig:hc1Overhc2Vshc3Overhc2}, as parametric
functions of the hole's spin $a$ and the orbital radius $r/r_{\rm
isco}$.  From this plot it is evident that the instantaneous amplitude 
ratios will give both $a$ and the instantaneous $r/r_{\rm isco}$ with 
moderate accuracy---though
only for those systems with strong enough signals that the weakest of
these harmonics, $m=1$, stands up strongly above the noise; cf.\ the
short-dashed curves in Figs.\
\ref{fig:hc1million}--\ref{fig:hc10tenmillion}.
  
In our idealized case of circular, equatorial orbits, this harmonic-ratio
information is not independent of that from the orbital phasing, but it 
could provide a confirmation of the phasing conclusions.  

In the more
realistic case of noncircular, nonequatorial orbits, the waveforms will
be much richer and there will be many more parameters to solve for.
Our survey of the circular, equatorial case gives some
rough indication of the kinds of information one can extract and by
what methods.

\section{Concluding Remarks}
\label{sec:Conclusions}

In this paper we have tabulated the results of TSN-based computations
of the waves emitted by an object spiraling into a spinning, massive
black hole on a slowly shrinking, circular, equatorial orbit.  Our
Tables \ref{ScrDotE.tbl}--\ref{ScrNorb.tbl} can serve as a foundation
for future mission-definition studies for LISA---most particularly, for
studies of how changes in the mission design may affect LISA's ability
to detect such inspiral waves, for studies of the accuracies with
which LISA's data can extract the properties of the source, and for
explorations of possible data analysis algorithms.

Much more important, in the long run, will be the extension of our
analysis to nonequatorial and noncircular orbits.  This extension is
urgent, since models of active galactic nuclei predict, rather firmly,
that the orbits will be nonequatorial and quite noncircular, and since
the earliest possible date for LISA to fly is less than ten years in
the future.

\section*{Acknowledgments}

For helpful discussions we thank Fintan Ryan and Eric Poisson. 
For information and advice about the LISA noise curve we thank John
Armstrong, Peter Bender, Curt Cutler, Frank Estabrook, Robin (Tuck) Stebbins,
and Massimo Tinto; and we thank Bender and 
Stebbins for providing us with a table
of the noise curve from Ref.\ \cite{mdt}.  For information and advice about
white-dwarf-binary background noise, we thank Peter Bender, Sterl Phinney   
and Tuck Stebbins.
This paper was supported in
part by NASA grants NAGW-4274, NAG5-6840 and their predecessors, 
and in view of its
future applications to LIGO, by NSF grants PHY-9800111, PHY-9996213,
AST-9731698 and their predecessors.

%

\appendix

\section*{Numerical methods}

Teukolsky \cite{teukolsky} found that the equations describing
perturbations of the Kerr spacetime could be separated into separate
radial and angular equations. For the circular, equatorial 
orbits studied in this paper, the challenges of solving the perturbation
equations are all associated with the numerical solution of the radial
equation.  We have used Green function methods to solve the radial
equation and determine the power radiated down the horizon and to
infinity by a particle in a circular equatorial orbit. The
general method of solution and formulation of the problem is well
described in Ref.\ \cite{hughes00a}, and we refer the interested reader there
for details. In this appendix we describe several innovations that can
dramatically speed the solution of the radial equation compared to the
more conventional methods applied elsewhere.

The Teukolsky radial equation is a second order, ordinary differential
equation. In the form given originally by Teukolsky
\cite{teukolsky} the equation is stiff and the solution satisfying
the physical boundary conditions is difficult to obtain.  Sasaki and
Nakamura \cite{sasaki_nakamura} found, through a local
change of variables, a form of the radial equation which is not stiff,
and we have worked with the radial equation in that form.

In the Sasaki-Nakamura formulation, the homogeneous (source-free)
radial equation takes the form
\begin{equation}\label{eq:sn}
    \left[{d\over dr_{*}} -{\cal F}(r_{*}){d\over dr_{*}} - {\cal
    U}(r_{*})\right]X = 0\;.
\end{equation}
Here $r_{*}$ is the so-called tortoise coordinate, which ranges from
$-\infty$ at the (outer) horizon to $\infty$ at spatial infinity. (In
this appendix, and in this appendix only, we express all dimensioned
quantities, such as $r$ and $r_{*}$, in terms of the black hole's mass
$M$, 
eschewing for convenience the superscript-tilde notation
used elsewhere in this paper.) The tortoise coordinate can be
expressed analytically in terms of the Boyer-Lindquist radial
coordinate $r$ and the location of the inner and outer horizons
$r_{+}$ and $r_{-}$:
\begin{mathletters}
\begin{eqnarray}
r_{*} = r + {2\over r_{+}-r_{-}}\left[
r_{+}\ln\left(r-r_{+}\right) -
r_{-}\ln\left(r-r_{-}\right)
\right]
\end{eqnarray}
where
\begin{eqnarray}
(r-r_{+})(r-r_{-}) &=& r^2-2r+a^2\qquad\text{and}\\
r_{+} &\geq& r_{-}\;.
\end{eqnarray}
\end{mathletters}
The functions ${\cal F}$ and ${\cal U}$ are parameterized by the
angular frequency of the perturbation $\omega = 2\pi f$, the angular momentum
of the spacetime $a$, and the angular separation constants $\ell$ and
$m$ (with $|m|\leq\ell$).  (For the particular forms of ${\cal F}$ and
${\cal U}$ see \cite{hughes00a}.)  For circular, equatorial orbits
$\omega$ is always an integer multiple of the orbital angular
frequency, $\omega= \omega_m \equiv m\Omega = 2\pi f_m$ .

To obtain the Green function solution to the radial equation with
source we need
the two solutions to the homogeneous equation corresponding to
the physical boundary conditions at infinity (no in-coming radiation)
and the horizon (no up-going radiation). These solutions are
determined numerically by posing the boundary conditions near infinity
or the horizon and integrating the radial equation inward or outward,
as appropriate.  In the Sasaki-Nakamura variables, obtaining a solution
to the radial equation poses no particular challenge; correspondingly,
it is conventional to use a ``work-horse'' integrator ({\em e.g.,}
Runge-Kutta or Bulirsch-Stoer) to solve the equation. On the other
hand, the radial equation arises from a separation of variables and is
parameterized by the separation constants $\ell$, $m$ and $\omega$,
corresponding to the resolved angular and temporal dependence of the
perturbation.  Consequently, it is necessary to solve the radial
equation separately for every important set of angular multipoles
($\ell$, $m$) and frequency $\omega$.  For very relativistic orbits
even moderate accuracy in the total radiated power may require solving
the radial equation tens of thousands of times for different angular
multipoles and harmonics of the orbital frequency.  Consequently,
speeding the solution while preserving its accuracy is of fundamental
importance.  In the remainder of this appendix we address several
innovations we have made in solving this equation that, depending on
the details of the orbit and the desired accuracy of the solution, can
result in a several order of magnitude reduction in the solution time
compared to a conventional approach.

\subsection{Boundary conditions at the horizon}

As one approaches the horizon, the physical solution for the radial
function, corresponding to down-going radiation, leads to the boundary
conditions used for the numerical integration of one of the
homogeneous solutions of the Sasaki-Nakamura equation:
\begin{mathletters}
    \label{eq:bc}
    \begin{eqnarray}
      \lim_{r_{*}\rightarrow-\infty} X_H(r_{*}) &=&
      e^{-i\omega_{-} r_{*}}\;,\\
      \lim_{r_{*}\rightarrow-\infty} {dX_H\over dr_{*}}(r_{*}) &=&
      -i\omega_{-}e^{-i\omega_{-} r_{*}}\;,
    \end{eqnarray}
    where
    \begin{eqnarray}
        \omega_{-} &=& \omega-{am\over2r_{+}}
    \end{eqnarray}
\end{mathletters}
and $r_{+}$ is the radius of the outer horizon in Boyer-Lindquist
coordinates. 

As a practical matter the boundary conditions used to determine $X_H$
are posed at some large, negative but finite $r_*$, say $R^{*}_{-}$;
{\em i.e.,} ``close to'', but not at, the horizon.  Using Eqs.\ 
(\ref{eq:bc}) evaluated at finite $R_{-}^{*}$ for the boundary
conditions introduces errors of fractional order $\delta = r-r_{+}$
into the solution.  This error can be represented as an error in the
amplitude of the power radiated down the horizon and the introduction
of some small component of radiation up-going from the horizon. These
errors propagate to large $r_{*}$ where they contribute to the
out-going radiation and lead to errors in the calculated power
radiated to infinity by the orbiting particle.

The errors introduced by using equations \ref{eq:bc} when posing
boundary conditions at 
finite radius can be expressed as a power series in $\delta$. The
coefficients of that expansion can be estimated by 
solving the equations several times, for different $R^{*}_{-}$, and
using Richardson extrapolation. To estimate the first $N$ terms in the
error expansion requires $N+1$ numerical solutions of the equations,
each beginning with the boundary conditions posed at a different
$R^{*}_{-}$. Controlling the error requires that the radial equation
be solved at least twice and often three or more times at different,
large $|R^{*}_{-}|$. 

To improve the convergence rate of this error estimate and allow us to
pose our boundary conditions at smaller $|R^{*}_{-}|$ we have solved the
Sasaki-Nakamura equation analytically
about the point at $r_{*}=-\infty$, finding the
first corrections in $\delta$ to the boundary equations given by
Eqs. (\ref{eq:bc}). The improved boundary conditions are given by 
\begin{mathletters}
    \label{eq:bc2}
    \begin{eqnarray}
    X(r_{*}) &=& 1 + \delta A'\;,\\
    {dX\over dr_{*}}(r_{*}) &=& 
    \left[
    -i\omega_{-}\left(1+\delta A'\right)
    +{\delta d\over 2r_{+}}A'
    \right]\;,
  \end{eqnarray}
  where
  \begin{eqnarray}
    A' &=&
    \left[
      {\cal F}^{(1)} +
      {d\over 4r_{+}^{2}}
      \left(U_{2}^{(1)} + F_{1}^{(1)} + G_{1}^{(1)}\right)
    \right]\nonumber\\
    &&\times
    \left[{d^{2}\over 4r_{+}^{2}}+i\epsilon {d\over 
        r_{+}}\omega_{-}\right]^{-1}\;,\\
    \delta &\equiv& r-r_{+}\;,\\
    d &\equiv& r_{+}-r_{-}\;,\\
    {\cal F}^{(1)} &\equiv& -i\omega_{-}{d\over2r_{+}}\;,\\
    U_{2}^{(1)} &\equiv& 
    \lambda + {4r_{-}\over r_{+}} -
    2r_{+}\omega_{-}
    \left(
      {2(2+d)\over d}
      - {2r_{+}\over d^{2}}\omega_{-}
    \right) \nonumber\\
    &&
    -\left(
      1 - {d(2+d)\over r_{+}}
    \right)
    {4r_{+}^{2}\over d^{2}}
    \omega_{-}^2\;,\\
    F^{(1)} &\equiv& 2ir_{+}\omega_{-}F(0)\;,\\
    G^{(1)}_{1} &\equiv& -2+d/2\;,\\
    F(0) &\equiv& \gamma'(r_{+})/\gamma(r_{+})\;,\\
    \gamma(r) &\equiv& \sum_{k=0}^4\gamma_k r^{-k}\;,\\
    \gamma_0 &\equiv& \lambda\left(\lambda+2\right) -
    12a\omega\left(a\omega-m\right) - 12 i \omega\;,\\
    \gamma_1 &\equiv& 8i\left[3a^2\omega -
      a\lambda\left(a\omega-m\right)\right]\;,\\ 
    \gamma_2 &\equiv& 12\left[ 
      -2ai\left(a\omega-m\right) + a^2 - 2a^2\left(a\omega-m\right)^2
    \right] \;,\\
    \gamma_3 &\equiv& 24a^2\left[-1+ia\left(a\omega-m\right)\right]\;,\\
    \gamma_4 &\equiv& 12a^4\;,\\
    \lambda &\equiv& \ell(\ell+1) - 2am\omega + a^2\omega^2 + 2\;.
    \end{eqnarray}
\end{mathletters}

The numerical solution to the radial equation using these improved
boundary conditions converges upon the true solution more quickly than
a solution using the boundary conditions (\ref{eq:bc}).  We are thus
able to pose approximate horizon boundary conditions at smaller 
$|R^{*}_{-}|$, reducing the domain over which we must integrate the radial
equation and, often the number of times we must integrate the equation
for each $(\omega,\ell,m)$ in order to obtain a solution of controlled
accuracy. 

\subsection{Boundary conditions at spatial infinity}

As $r_{*}\rightarrow\infty$, the physical solution for the radial
function, corresponding to no in-going radiation, leads to the
boundary conditions for the numerical integration of the other
critical solution of the radial equation: 
\begin{eqnarray}
  \lim_{r_{*}\rightarrow\infty} X_\infty &=& 1\;,\\
  \lim_{r_{*}\rightarrow\infty} {dX_\infty\over dr_{*}} &=& i\omega\; . 
\end{eqnarray}

As with the boundary conditions at the horizon, we construct the
solution $X_\infty$ beginning with boundary conditions posed at finite
$R^{*}_{+}$, not at infinity. Using the asymptotic form of the
boundary conditions to set $X$ and $X'$ at finite radius leads to
errors of fractional order $1/R^{*}_{+}$ in the solution, which can be
represented as an error in the amplitude of the out-going radiation
and the introduction of some small in-going radiation component. These
lead, in turn, to errors in the estimated power radiated to infinity
and down the horizon.  We can use Richardson extrapolation to estimate
and reduce this error; however, as before, the radiated power must be
determined at several different large $R_{+}^{*}$ in order to estimate
and reduce the error.

To permit a more accurate estimate of the radiated power from $X$ and
$X'$ evaluated at smaller $R_{+}^{*}$ we have solved the
Sasaki-Nakamura equation analytically 
about the point at $r_{*}=\infty$, finding
the first corrections in $1/R^{*}_{+}$ to the asymptotic form of the
radial function $X$. For finite $R_{+}^{*}$ we have
\begin{mathletters}
\begin{eqnarray}
X_\infty(r_{*}) &=& 
\left(1+{a_1\over r}\right)\;,\\
{dX_\infty\over dr_{*}}(r_{*}) &=&
i\omega 
\left(1+{a_1\over r}\right)\;,
\end{eqnarray}
where 
\begin{eqnarray}
a_1 &=& {\gamma_{1}\over\gamma_0} +
{i\over2} \left[
  \omega\left(a^2+4\right) + 2am + {\lambda+2\over\omega}
\right]\;.
\end{eqnarray}
\end{mathletters}
We use these expressions, evaluated at finite but large $R^{*}_{+}$,
to set the boundary condition for the numerical solution of the
homogeneous radial equation. We continue to use Richardson
extrapolation to control the error of the solutions; however, each
step in the extrapolation has a greater effect on the error and the 
extrapolation can take place at smaller $R^{*}_{+}$. 

\subsection{A more suitable choice of variables}

The solution $X$ to the Sasaki-Nakamura equations is a complex
oscillatory function.  Integrating the equations directly for $X$
requires a spatial resolution $\Delta r_{*}$ less than the local
wavelength of $X$, 
\begin{equation}
    \Delta r_{*}\lesssim \left|d\ln X\over dr_{*}\right|^{-1}.
\end{equation}
When solving for the radial function corresponding to a high temporal
frequency $|\omega|$ the step-size can become quite small, with a
corresponding increase in the computational time for an accurate
solution.

It is advantageous in circumstances like these to reformulate the 
problem in action-angle variables, whose variation is both slower and 
smoother than the variations in $X$. Writing $X$ as 
\begin{mathletters}
\begin{equation}
    X \equiv \exp\left[i\Phi(r_{*})\right]
\end{equation}
we define the two real functions $\xi$ and $\phi$ as the 
imaginary and real parts of $\Phi$:
\begin{eqnarray}
    \xi &\equiv& \Re\left(\Phi\right)\\
    \phi &\equiv& \Im\left(\Phi\right)
\end{eqnarray}
\end{mathletters}
With this substitution the linear Sasaki-Nakamura equation for complex
$X$ becomes a pair of coupled non-linear equations for the real $\xi$
and $\phi$.  The equation for $\xi$ is second order while the equation
for $\phi$ can be integrated immediately to obtain a first order
equation.  (This is expected since the solution for $X$ is determined
only up to an overall phase.)  Both $\phi$ and $\xi$ vary slowly and
smoothly compared to $X$.  This is particularly true as one moves
toward either the horizon or spatial infinity, where $X$ is
oscillatory in $r_{*}$ while $\xi$ is constant and $\phi$ is linear.
Correspondingly, the numerical solution of the equations for $\phi$
and $\xi$ require much less resolution for the same numerical
accuracy, dramatically speeding the integration of the radial
equation.

\subsection{Numerical solution of the equations for $\xi$ and $\phi$}

The local errors committed by, {\em e.g.,} a fourth order Runge-Kutta
integration of the radial equation are proportional to $\Delta
{r_{*}}^{5}$.  Reducing the step-size and increasing the number of
integration steps will decrease the overall solution error
algebraically, {\em i.e.,} as a fixed power of $\Delta r_{*}$, while
increasing the time required for a solution. A higher order
computational method will increase the solution accuracy more rapidly.
{\em Exponential\/} convergence of the solution with $\Delta r_{*}$
can be obtained if the equations are solved via collocation
pseudo-spectral techniques \cite{gottleib77a}.  In a collocation
pseudo-spectral method the solution for the dependent variable is
approximated as a sum over a suitable set of basis functions.  The
differential equations, evaluated on the approximate solution at a
fixed number of points, then determine the coefficients in the
expansion.  For problems with smooth solutions the solution accuracy
increases exponentially with the number of terms in the approximation
(and, correspondingly, with the number of evaluations of the
differential equation, which is the analog of the spatial resolution
of the integration).  Our final innovation is to solve the radial
equation using pseudo-spectral techniques.  We have chosen a
Chebyshev expansion for $\xi$ and $\phi$ with Gauss-Lobatto
collocation points.  Our experience is that the best performance is
obtain if the integration domain $[R_{+},R_{-}]$ is divided into two
parts, at approximately the peak of the effective potential $R_{0}$:
{\em i.e.,} we use two expansions for $\phi$ and $\xi$, one in the
domain $[R_{+},R_{0}]$ and the other in the domain $[R_{0},R_{-}]$.
At $R_{0}$ we insist that the two solutions for $\phi$ and $\xi$ agree
in value, and that the solutions for $\xi$ agree also in their first
derivative, as is appropriate for functions described by first order
and second order differential equations, respectively.

\newpage
\widetext
\squeezetable
\begin{table}
\caption{$\Omega / \Omega_{\rm isco}$ (orbital angular velocity in units of
that at the isco) as a function of $r/r_{\rm isco}$ (Boyer-Lindquist 
radius in units of that at the isco)
and of $a$ (black-hole angular momentum parameter).  For $a$ negative, 
the hole is counter-rotating relative to the star's orbit; for $a$ positive it
is co-rotating.  This table was computed from Eqs.\ (\protect\ref{rOmega}), 
(\protect\ref{Omegams}) and (\protect\ref{rms}).  Near the isco, $\Omega /
\Omega{\rm isco}$ is linear in $r/r_{\rm isco}$.
}
\begin{tabular}{d|dddddddddd}
$r/r_{\rm isco}$  &     $-$0.99 &     $-$0.9 &     $-$0.5 &      0.0 &      0.2 &      0.5 &      0.8 &      0.9 &      0.99 &      0.999 \\\tableline
1.000& 1.0& 1.0& 1.0& 1.0& 1.0& 1.0& 1.0& 1.0& 1.0& 1.0\\
1.001& 0.9984& 0.9984& 0.9985& 0.9985& 0.9985& 0.9986& 0.9987&     0.9988& 0.9990& 0.9992\\
1.002& 0.9969& 0.9969& 0.9969& 0.9970&     0.9971& 0.9972& 0.9974& 0.9976& 0.9981& 0.9983\\
1.005& 0.9923& 0.9923& 0.9924& 0.9925& 0.9927& 0.9929& 0.9936&     0.9940& 0.9952& 0.9958\\
1.01& 0.9846& 0.9847& 0.9848& 0.9852&     0.9854& 0.9860& 0.9872& 0.9882& 0.9905& 0.9916\\
1.02& 0.9696& 0.9697& 0.97& 0.9707& 0.9712& 0.9723& 0.9747&     0.9765& 0.9811& 0.9833\\
1.05& 0.9269& 0.9271& 0.9278& 0.9294&     0.9305& 0.9330& 0.9386& 0.9429& 0.9537& 0.9590\\
1.1& 0.8624& 0.8626& 0.8639& 0.8668& 0.8686& 0.8731& 0.8831&     0.8909& 0.9105& 0.9204\\
1.2& 0.7538& 0.7542& 0.7563& 0.7607&     0.7636& 0.7707& 0.7869& 0.7995& 0.8326& 0.8497\\
1.3& 0.6664& 0.6668& 0.6693& 0.6747& 0.6782& 0.6868& 0.7066&     0.7223& 0.7644& 0.7866\\
1.4& 0.5947& 0.5951& 0.5978& 0.6037&     0.6075& 0.6170& 0.6389& 0.6565& 0.7044& 0.7303\\
1.7& 0.4419& 0.4424& 0.4451& 0.4512& 0.4552& 0.4650& 0.4884&     0.5077& 0.5625& 0.5937\\
2.0& 0.3450& 0.3455& 0.3480& 0.3536&     0.3572& 0.3664& 0.3885& 0.4069& 0.4611& 0.4930\\
2.5& 0.2460& 0.2463& 0.2484& 0.2530& 0.2560& 0.2637& 0.2823& 0.2982&     0.3463& 0.3758\\
3.0& 0.1867& 0.1870& 0.1887& 0.1925& 0.1950&     0.2013& 0.2168& 0.2302& 0.2716& 0.2976\\
4.0& 0.1210& 0.1212& 0.1224& 0.1250& 0.1268& 0.1312& 0.1423& 0.1520&     0.1827& 0.2025\\
5.0& 0.08643& 0.08659& 0.08748& 0.08944& 0.09076&     0.09409& 0.1024& 0.1097& 0.1332& 0.1487\\
6.0& 0.06570& 0.06582& 0.06651& 0.06804& 0.06907& 0.07167& 0.07817&     0.08391& 0.1025& 0.1149\\
7.0& 0.05211& 0.05221& 0.05276& 0.05399& 0.05482& 0.05692& 0.06217&     0.06682& 0.08197& 0.09211\\
8.0& 0.04264& 0.04271& 0.04318& 0.04419& 0.04488& 0.04661& 0.05097&     0.05483& 0.06745& 0.07595\\
9.0& 0.03572& 0.03579& 0.03618& 0.03704& 0.03762& 0.03908& 0.04276&     0.04603& 0.05675& 0.06399\\
10.0& 0.03049& 0.03055& 0.03088& 0.03162& 0.03212& 0.03338&     0.03654& 0.03936& 0.04860& 0.05486
\end{tabular}
\label{Omega.tbl}
\end{table}

\begin{table}
\caption{$\dot{\cal E}$ (the relativistic correction to $\dot E_{\rm GW} = -
\dot E$, the total rate of emission of energy into gravitational waves 
going both to infinity and down the hole), as a function of orbital radius
$r/r_{\rm isco}$ and black-hole spin parameter $a$; cf.\ caption of Table
\protect\ref{Omega.tbl}.  
This table is accurate to four significant digits; each entry 
was computed by summing over enough spheroidal
harmonic orders $(l,m)$ to produce that accuracy. 
}
\begin{tabular}{d|dddddddddd}
$r/r_{\rm isco}$  &     $-$0.99 &     $-$0.9 &     $-$0.5 &      0.0 &      0.2 &      0.5 &      0.8 &      0.9 &      0.99 &      0.999 \\\tableline
1.000& 1.240& 1.233& 1.197& 1.143& 1.114& 1.053& 0.9144& 0.7895&     0.4148& 0.2022\\ 
1.001& 1.239& 1.232& 1.196& 1.142& 1.114&     1.053& 0.9140& 0.7894& 0.4154& 0.2032\\
1.002& 1.238& 1.231& 1.196& 1.141& 1.113& 1.052& 0.9137& 0.7894&     0.4160& 0.2041\\
1.005& 1.235& 1.228& 1.193& 1.139& 1.111& 1.050&     0.9126& 0.7891& 0.4177& 0.2069\\
1.01& 1.231& 1.224& 1.189& 1.135& 1.107& 1.047& 0.9109& 0.7887&     0.4207& 0.2116\\
1.02& 1.222& 1.215& 1.181& 1.127& 1.100& 1.041&     0.9076& 0.7880& 0.4263& 0.2208\\
1.05& 1.198& 1.192& 1.159& 1.108& 1.081& 1.025& 0.8988& 0.7867&     0.4434& 0.2473\\
1.1& 1.165& 1.159& 1.128& 1.080& 1.055& 1.002&     0.8876& 0.7859& 0.4701& 0.2881\\
1.2& 1.115& 1.110& 1.082& 1.039& 1.017& 0.9706& 0.8726& 0.7882&     0.5182& 0.3581\\
1.3& 1.081& 1.075& 1.051& 1.012& 0.9913&     0.9493& 0.8638& 0.7920& 0.5587& 0.4160\\
1.4& 1.055& 1.051& 1.028& 0.9919& 0.9733& 0.9348& 0.8583& 0.7960&     0.5930& 0.4648\\
1.7& 1.011& 1.007& 0.9888& 0.9591& 0.9435&     0.9119& 0.8524& 0.8075& 0.6665& 0.5723\\
2.0& 0.9893& 0.9862& 0.9705& 0.9448& 0.9312& 0.9034& 0.8530&     0.8171& 0.7117& 0.6411\\
2.5& 0.9734& 0.9709& 0.9580& 0.9363&     0.9248& 0.9012& 0.8589& 0.8302& 0.7556& 0.7089\\
3.0& 0.9674& 0.9653& 0.9542& 0.9352& 0.9250& 0.9040& 0.8662& 0.8415&     0.7813& 0.7469\\
4.0& 0.9651& 0.9634& 0.9546& 0.9391& 0.9306&     0.9129& 0.8807& 0.8597& 0.8121& 0.7882\\
5.0& 0.9665& 0.9651& 0.9577& 0.9448& 0.9371& 0.9216& 0.8930&     0.8742& 0.8320& 0.8118\\
6.0& 0.9687& 0.9675& 0.9611& 0.9490&     0.9430& 0.9291& 0.9031& 0.8858& 0.8469& 0.8286\\
7.0& 0.9709& 0.9699& 0.9641& 0.9533& 0.9480& 0.9354& 0.9116&     0.8955& 0.8589& 0.8416\\
8.0& 0.9730& 0.9720& 0.9669& 0.9588&     0.9522& 0.9407& 0.9186& 0.9036& 0.8689& 0.8524\\
9.0& 0.9749& 0.9740& 0.9693& 0.9607& 0.9558& 0.9452& 0.9246&     0.9105& 0.8774& 0.8616\\
10.0& 0.9765& 0.9757& 0.9714& 0.9616&     0.9589& 0.9491& 0.9298& 0.9164& 0.8847& 0.8695\\
\end{tabular}
\label{ScrDotE.tbl}
\end{table}

\newpage
\null
\newpage
\begin{table}
\caption{$\dot{\cal E}_{\infty 1}$ 
(the relativistic correction to $\dot E_{\infty 1}$, the rate of emission of
energy into harmonic-1 gravitational waves with frequency $f_1 = \Omega/2\pi$
traveling to infinity) as a function of orbital radius
$r/r_{\rm isco}$ and black-hole spin parameter $a$; cf.\ caption of Table
\protect\ref{Omega.tbl}.  
This table is accurate to four significant digits; each entry
was computed by summing over enough spheroidal
harmonic orders $2\le l \le l_{\rm max}$ at fixed $|m|=1$ to produce that
accuracy.
}

\begin{tabular}{d|dddddddddd}
$r/r_{\rm isco}$  &     $-$0.99 &     $-$0.9 &     $-$0.5 &      0.0 &      0.2 &      0.5 &      0.8 &      0.9 &      0.99 &      0.999 \\\tableline
1.000& 3.013& 2.854& 2.157& 1.320& 1.002& 0.5530& 0.1669& 0.06573&     0.002762& 1.071E-4\\
1.001& 3.010& 2.851& 2.156& 1.319& 1.001&     0.5529& 0.1670& 0.06584& 0.002783& 1.095E-4\\
1.002& 3.007& 2.849& 2.154& 1.318& 1.001& 0.5528& 0.1671&     0.06595& 0.002805& 1.119E-4\\
1.005& 2.998& 2.840& 2.148& 1.316& 0.9990& 0.5525& 0.1675& 0.06628&     0.002869& 1.194E-4\\ 
1.01& 2.984& 2.827& 2.139& 1.312& 0.9964&     0.5520& 0.1680& 0.06683& 0.002979& 1.326E-4\\    
1.02& 2.955& 2.800& 2.121& 1.303& 0.9915& 0.5510& 0.1692& 0.06793&     0.003204& 1.619E-4\\ 
1.05& 2.876& 2.727& 2.071& 1.280& 0.9779&     0.5487& 0.1729& 0.07126& 0.003934& 2.747E-4\\    
1.1& 2.760& 2.619& 1.999& 1.248& 0.9588& 0.5462& 0.1790& 0.07688&     0.005322& 5.565E-4\\ 
1.2& 2.575& 2.448& 1.885& 1.198& 0.9305&     0.5448& 0.1916& 0.08821& 0.008679& 0.001525\\    
1.3& 2.434& 2.316& 1.798& 1.161& 0.9111& 0.5465& 0.2040& 0.09951&     0.01272& 0.003073\\ 
1.4& 2.321& 2.213& 1.730& 1.133& 0.8973&     0.5498& 0.2162& 0.1107& 0.01733& 0.005180\\    
1.7& 2.085& 1.994& 1.588& 1.077& 0.8719& 0.5623& 0.2506& 0.1421&     0.03343& 0.01435\\ 
2.0& 1.940& 1.860& 1.503& 1.049& 0.8631& 0.5786&     0.2799& 0.1715& 0.05138& 0.02648\\    
2.5& 1.787& 1.719& 1.416& 1.023& 0.8599& 0.6045& 0.3237& 0.2151&     0.08223& 0.04991\\ 
3.0& 1.689& 1.629& 1.361& 1.010& 0.8621&     0.6272& 0.3606& 0.2527& 0.1122& 0.07456\\    
4.0& 1.567& 1.518& 1.295& 0.9987& 0.8704& 0.6638& 0.4191& 0.3143&     0.1664& 0.1222\\ 
5.0& 1.493& 1.450& 1.255& 0.9921& 0.8789& 0.6918&     0.4638& 0.3627& 0.2128& 0.1650\\    
6.0& 1.441& 1.403& 1.228& 0.9923& 0.8865& 0.7139& 0.4994& 0.4020&     0.2525& 0.2028\\ 
7.0& 1.403& 1.368& 1.208& 0.9865& 0.8930& 0.7319&     0.5287& 0.4346& 0.2869& 0.2362\\    
8.0& 1.373& 1.341& 1.193& 0.9829& 0.8987& 0.7469& 0.5533& 0.4624&     0.3168& 0.2657\\ 
9.0& 1.349& 1.319& 1.180& 0.9887& 0.9035& 0.7596&     0.5743& 0.4863& 0.3433& 0.2922\\    
10.0& 1.329& 1.301& 1.170& 1.005& 0.9078& 0.7706& 0.5925& 0.5072&     0.3669& 0.3159\\

\end{tabular}
\label{ScrDotE1.tbl}
\end{table}

\begin{table}
\caption{$\dot{\cal E}_{\infty 2}$
(the relativistic correction to $\dot E_{\infty 2}$, the rate of emission of
energy into harmonic-2 gravitational waves with frequency $f_2 = 2 \Omega/2\pi$
traveling to infinity) as a function of orbital radius
$r/r_{\rm isco}$ and black-hole spin parameter $a$; cf.\ caption of Table
\protect\ref{Omega.tbl}. 
This table is accurate to four significant digits; each entry
was computed by summing over enough spheroidal
harmonic orders $2\le l \le l_{\rm max}$ at fixed $|m|=2$ to produce that
accuracy.
}
\begin{tabular}{d|dddddddddd}
$r/r_{\rm isco}$  &     $-$0.99 &     $-$0.9 &     $-$0.5 &      0.0 &      0.2 &      0.5 &      0.8 &      0.9 &      0.99 &      0.999 \\\tableline
1.000& 1.029& 1.020& 0.9734& 0.8957& 0.8535& 0.7653& 0.5914& 0.4617&     0.1656& 0.06128\\ 
1.001& 1.028& 1.019& 0.9730& 0.8954& 0.8533&     0.7652& 0.5915& 0.4620& 0.1661& 0.06170\\    
1.002& 1.028& 1.019& 0.9726& 0.8950& 0.8530& 0.7650& 0.5916& 0.4624&     0.1666& 0.06212\\ 
1.005& 1.026& 1.017& 0.9713& 0.8940& 0.8522&     0.7645& 0.5919& 0.4633& 0.1681& 0.06338\\    
1.01& 1.024& 1.015& 0.9693& 0.8925& 0.8508& 0.7638& 0.5925&     0.4649& 0.1707& 0.0655\\ 
1.02& 1.019& 1.011& 0.9654& 0.8894&     0.8483& 0.7623& 0.5937& 0.4680& 0.1758& 0.06975\\    
1.05& 1.007& 0.9985& 0.9548& 0.8813& 0.8415& 0.7587& 0.5974&     0.4773& 0.1909& 0.08241\\    
1.1& 0.9900& 0.9818& 0.9403& 0.8704& 0.8327& 0.7545& 0.6037&     0.4918& 0.2154& 0.1038\\ 
1.2& 0.9648& 0.9574& 0.9196& 0.8558&     0.8214& 0.7506& 0.6165& 0.5178& 0.2618& 0.1465\\    
1.3& 0.9480& 0.9411& 0.9063& 0.8474& 0.8156& 0.7504& 0.6287&     0.5402& 0.3039& 0.1881\\ 
1.4& 0.9364& 0.9301& 0.8977& 0.8427&     0.8131& 0.7523& 0.6400& 0.5597& 0.3417& 0.2277\\    
1.7& 0.9191& 0.9138& 0.8867& 0.8403& 0.8151& 0.7635& 0.6697&     0.6050& 0.4309& 0.3303\\ 
2.0& 0.9138& 0.9092& 0.8857& 0.8450&     0.8227& 0.7769& 0.6941& 0.6382& 0.4930& 0.4077\\    
2.5& 0.9144& 0.9106& 0.8910& 0.8566& 0.8377& 0.7983& 0.7268&     0.6791& 0.5617& 0.4954\\ 
3.0& 0.9187& 0.9154& 0.8984& 0.8684&     0.8517& 0.8167& 0.7525& 0.7097& 0.6075& 0.5524\\    
4.0& 0.9286& 0.9260& 0.9125& 0.8882& 0.8745& 0.8453& 0.7908&     0.7542& 0.6680& 0.6241\\ 
5.0& 0.9372& 0.9351& 0.9237& 0.9034&     0.8914& 0.8662& 0.8183& 0.7857& 0.7085& 0.6699\\    
6.0& 0.9442& 0.9424& 0.9326& 0.9142& 0.9043& 0.8821& 0.8391&     0.8095& 0.7387& 0.7033\\ 
7.0& 0.9499& 0.9483& 0.9396& 0.9232&     0.9145& 0.8945& 0.8554& 0.8282& 0.7625& 0.7295\\    
8.0& 0.9545& 0.9531& 0.9453& 0.9323& 0.9226& 0.9044& 0.8686&     0.8434& 0.7819& 0.7507\\ 
9.0& 0.9584& 0.9571& 0.9500& 0.9369&     0.9293& 0.9126& 0.8795& 0.8560& 0.7981& 0.7685\\    
10.0& 0.9616& 0.9604& 0.9540& 0.9400& 0.9349& 0.9195& 0.8887& 0.8666&     0.8119& 0.7837\\
\end{tabular}
\label{ScrDotE2.tbl}
\end{table}

\newpage
\null
\newpage

\begin{table}
\caption{$\dot{\cal E}_{\infty 3}$
(the relativistic correction to $\dot E_{\infty 3}$, the rate of emission of
energy into harmonic-3 gravitational waves with frequency $f_3 = 3 \Omega/2\pi$
traveling to infinity) as a function of orbital radius
$r/r_{\rm isco}$ and black-hole spin parameter $a$; cf.\ caption of Table
\protect\ref{Omega.tbl}. 
This table is accurate to four significant digits; each entry
was computed by summing over enough spheroidal
harmonic orders $2\le l \le l_{\rm max}$ at fixed $|m|=3$ to produce that
accuracy.
}
\begin{tabular}{d|dddddddddd}
$r/r_{\rm isco}$  &     $-$0.99 &     $-$0.9 &     $-$0.5 &      0.0 &      0.2 &      0.5 &      0.8 &      0.9 &      0.99 &      0.999 \\\tableline
1.000& 0.9753& 0.9614& 0.8926& 0.7848& 0.7309& 0.6292& 0.4684&     0.3712& 0.1573& 0.06456\\    
1.001& 0.9748& 0.9608& 0.8922& 0.7845& 0.7307& 0.6291& 0.4685&     0.3714& 0.1577& 0.06495\\    
1.002& 0.9742& 0.9603& 0.8917& 0.7842& 0.7304& 0.6289& 0.4685&     0.3715& 0.1581& 0.06534\\    
1.005& 0.9726& 0.9587& 0.8904& 0.7832& 0.7296& 0.6285& 0.4685&     0.3719& 0.1592& 0.06651\\    
1.01& 0.9699& 0.9561& 0.8882& 0.7817& 0.7284& 0.6277& 0.4686&     0.3725& 0.1610& 0.06846\\ 
1.02& 0.9648& 0.9512& 0.8841& 0.7787&     0.7260& 0.6263& 0.4688& 0.3739& 0.1646& 0.07237\\    
1.05& 0.9507& 0.9376& 0.8729& 0.7708& 0.7197& 0.6229& 0.4699&     0.3780& 0.1751& 0.08391\\ 
1.1& 0.9313& 0.9189& 0.8576& 0.7606&     0.7118& 0.6191& 0.4728& 0.3852& 0.1919& 0.1026\\    
1.2& 0.9033& 0.8921& 0.8365& 0.7476& 0.7026& 0.6166& 0.4806&     0.3998& 0.2223& 0.1374\\ 
1.3& 0.8849& 0.8747& 0.8235& 0.7411&     0.6990& 0.6183& 0.4900& 0.4141& 0.2489& 0.1686\\    
1.4& 0.8728& 0.8633& 0.8157& 0.7385& 0.6988& 0.6223& 0.5000& 0.4278&     0.2722& 0.1963\\ 
1.7& 0.8562& 0.8483& 0.8085& 0.7424& 0.7078&     0.6402& 0.5301& 0.4649& 0.3280& 0.2627\\    
2.0& 0.8534& 0.8466& 0.8117& 0.7531& 0.7221& 0.6603& 0.5580&     0.4969& 0.3701& 0.3114\\ 
2.5& 0.8588& 0.8531& 0.8239& 0.7737&     0.7466& 0.6919& 0.5986& 0.5417& 0.4238& 0.3710\\    
3.0& 0.8676& 0.8627& 0.8373& 0.7929& 0.7687& 0.7190& 0.6324&     0.5785& 0.4657& 0.4156\\ 
4.0& 0.8851& 0.8811& 0.8606& 0.8243&     0.8040& 0.7616& 0.6850& 0.6358& 0.5297& 0.4822\\    
5.0& 0.8993& 0.8961& 0.8788& 0.8477& 0.8301& 0.7929& 0.7240&     0.6786& 0.5780& 0.5320\\ 
6.0& 0.9106& 0.9078& 0.8928& 0.8658& 0.8500&     0.8168& 0.7541& 0.7119& 0.6164& 0.5717\\    
7.0& 0.9198& 0.9172& 0.9039& 0.8787& 0.8657& 0.8356& 0.7780&     0.7387& 0.6478& 0.6045\\ 
8.0& 0.9272& 0.9249& 0.9130& 0.8902&     0.8783& 0.8508& 0.7975& 0.7607& 0.6741& 0.6322\\    
9.0& 0.9334& 0.9313& 0.9204& 0.9004& 0.8887& 0.8633& 0.8137&     0.7791& 0.6965& 0.6559\\ 
10.0& 0.9386& 0.9367& 0.9267& 0.9087&     0.8974& 0.8739& 0.8275& 0.7948& 0.7158& 0.6766\\
\end{tabular}
\label{ScrDotE3.tbl}
\end{table}

\begin{table}
\caption{$\dot{\cal E}_{\infty 4}$
(the relativistic correction to $\dot E_{\infty 4}$, the rate of emission of
energy into harmonic-4 gravitational waves with frequency $f_4 = 4 \Omega/2\pi$
traveling to infinity) as a function of orbital radius
$r/r_{\rm isco}$ and black-hole spin parameter $a$; cf.\ caption of Table
\protect\ref{Omega.tbl}. 
This table is accurate to four significant digits; each entry
was computed by summing over enough spheroidal
harmonic orders $2\le l \le l_{\rm max}$ at fixed $|m|=4$ to produce that
accuracy.
} 
\begin{tabular}{d|dddddddddd}
$r/r_{\rm isco}$  &     $-$0.99 &     $-$0.9 &     $-$0.5 &      0.0 &      0.2 &      0.5 &      0.8 &      0.9 &      0.99 &      0.999 \\\tableline
1.000& 0.9393& 0.9209& 0.8319& 0.6981& 0.6342& 0.5196& 0.3574& 0.2720&     0.1116& 0.04673\\ 
1.001& 0.9387& 0.9203& 0.8314& 0.6978& 0.6339&     0.5194& 0.3574& 0.2721& 0.1118& 0.04699\\    
1.002& 0.9380& 0.9197& 0.8309& 0.6975& 0.6337& 0.5193& 0.3575&     0.2722& 0.1120& 0.04726\\ 
1.005& 0.9361& 0.9179& 0.8295& 0.6965&     0.6329& 0.5189& 0.3575& 0.2725& 0.1127& 0.04807\\    
1.01& 0.9330& 0.9149& 0.8271& 0.6949& 0.6317& 0.5183& 0.3576&     0.2730& 0.1139& 0.04940\\ 
1.02& 0.9270& 0.9091& 0.8225& 0.6920&     0.6295& 0.5171& 0.3578& 0.2739& 0.1163& 0.05207\\    
1.05& 0.9106& 0.8935& 0.8102& 0.6841& 0.6235& 0.5143& 0.3589&     0.2770& 0.1231& 0.05993\\ 
1.1& 0.8881& 0.8720& 0.7936& 0.6740&     0.6162& 0.5115& 0.3617& 0.2826& 0.1342& 0.07255\\    
1.2& 0.8560& 0.8416& 0.7709& 0.6618& 0.6084& 0.5109& 0.3697&     0.2947& 0.1547& 0.09588\\    
1.3& 0.8353& 0.8221& 0.7573& 0.6563& 0.6064& 0.5143& 0.3793&     0.3073& 0.1731& 0.1168\\ 
1.4& 0.8218& 0.8097& 0.7496& 0.6549&     0.6077& 0.5199& 0.3897& 0.3198& 0.1898& 0.1355\\    
1.7& 0.8035& 0.7943& 0.7440& 0.6627& 0.6212& 0.5424& 0.4217&     0.3555& 0.2325& 0.1821\\ 
2.0& 0.8021& 0.7934& 0.7493& 0.6774&     0.6398& 0.5669& 0.4523& 0.3880& 0.2676& 0.2190\\    
2.5& 0.8109& 0.8036& 0.7664& 0.7037& 0.6710& 0.6052& 0.4977&     0.4355& 0.3163& 0.2683\\ 
3.0& 0.8232& 0.8169& 0.7843& 0.7286&     0.6984& 0.6383& 0.5365& 0.4759& 0.3570& 0.3086\\    
4.0& 0.8466& 0.8416& 0.8151& 0.7687& 0.7432& 0.6905& 0.5983&     0.5409& 0.4233& 0.3737\\ 
5.0& 0.8656& 0.8614& 0.8389& 0.7987&     0.7765& 0.7297& 0.6450& 0.5910& 0.4758& 0.4257\\    
6.0& 0.8807& 0.8770& 0.8574& 0.8227& 0.8021& 0.7599& 0.6818&     0.6306& 0.5190& 0.4689\\ 
7.0& 0.8928& 0.8895& 0.8720& 0.8412&     0.8224& 0.7838& 0.7115& 0.6631& 0.5551& 0.5055\\    
8.0& 0.9026& 0.8997& 0.8839& 0.8546& 0.8388& 0.8033& 0.7359&     0.6902& 0.5859& 0.5371\\ 
9.0& 0.9108& 0.9081& 0.8938& 0.8673&     0.8523& 0.8195& 0.7563& 0.7130& 0.6125& 0.5647\\    
10.0& 0.9177& 0.9152& 0.9020& 0.8759& 0.8637& 0.8331& 0.7737&     0.7326& 0.6355& 0.5889\\
\end{tabular}
\label{ScrDotE4.tbl}
\end{table}

\newpage
\null
\newpage

\begin{table}
\caption{
$\dot E_{\rm H}/\dot E_{\rm GW}$ (the ratio of the energy radiated down the
hole to the total energy radiated) 
as a function of orbital radius
$r/r_{\rm isco}$ and black-hole spin parameter $a$; cf.\ caption of Table
\protect\ref{Omega.tbl}. 
This table is accurate to three significant digits; each entry
was computed by summing over enough spheroidal
harmonic orders $(l,m)$ to produce that accuracy.
} 

\begin{tabular}{d|dddddddddd}
$r/r_{\rm ms}$  &     $-$0.99 &     $-$0.9 &     $-$0.5 &      0.0 &      0.2 &      0.5 &      0.8 &      0.9 &      0.99 &      0.999 \\\tableline
1.000& 0.0129& 0.0118& 0.00757& 0.00319& 0.00162& -0.00222&     -0.0166& -0.0341& -0.0942& -0.129\\    
1.001& 0.0129& 0.0118& 0.00753& 0.00318& 0.00161& -0.00222&     -0.0165& -0.0341& -0.0942& -0.129\\    
1.002& 0.0128& 0.0117& 0.00750& 0.00316& 0.00160& -0.00222&     -0.0165& -0.0341& -0.0942& -0.129\\    
1.005& 0.0127& 0.0116& 0.00740& 0.00310& 0.00156& -0.00224&     -0.0165& -0.0340& -0.0941& -0.129\\    
1.01& 0.0124& 0.0114& 0.00723& 0.00301& 0.00149& -0.00225&     -0.0164& -0.0339& -0.0941& -0.129\\    
1.02& 0.0120& 0.0109& 0.00691& 0.00284& 0.00137& -0.00228&     -0.0163& -0.0337& -0.0939& -0.129\\    
1.05& 0.0107& 0.00975& 0.00606& 0.00239& 0.00106& -0.00233&     -0.0159& -0.0330& -0.0930& -0.128\\    
1.1& 0.00898& 0.00814& 0.00493& 0.00182& 6.92E-4& -0.00234&     -0.0151& -0.0316& -0.0906& -0.125\\    
1.2& 0.00651& 0.00586& 0.00339& 0.00111& 2.72E-4& -0.00219&     -0.0137& -0.0285& -0.0834& -0.116\\    
1.3& 0.00489& 0.00438& 0.00244& 7.21E-4& 7.09E-5& -0.00198&     -0.0122& -0.0254& -0.0764& -0.107\\    
1.4& 0.00378& 0.00336& 0.00182& 4.89E-4& -2.80E-5& -0.00177&     -0.0109& -0.0227& -0.0692& -0.0980\\    
1.7& 0.00198& 0.00174& 8.80E-4& 1.85E-4& -1.08E-4& -0.00125&     -0.00781& -0.0164& -0.0519& -0.0754\\    
2.0& 0.00118& 0.00103& 5.00E-4& 8.52E-5& -1.01E-4& -9.03E-4&     -0.00570& -0.0123& -0.0396& -0.0586\\    
2.5& 6.03E-4& 5.22E-4& 2.41E-4& 3.07E-5& -7.20E-5& -5.60E-4&     -0.00359& -0.00792& -0.0264& -0.0401\\    
3.0& 3.55E-4& 3.05E-4& 1.37E-4& 1.37E-5& -5.01E-5& -3.71E-4&     -0.00241& -0.00539& -0.0186& -0.0287\\    
4.0& 1.58E-4& 1.35E-4& 5.88E-5& 3.94E-6& -2.64E-5& -1.89E-4&     -0.00125& -0.00284& -0.0103& -0.0164\\    
5.0& 8.62E-5& 7.34E-5& 3.13E-5& 1.530E-6& -1.55E-5& -1.10E-4& -7.41E-4&     -0.00169& -0.00637& -0.0103\\    
6.0& 5.29E-5& 4.50E-5& 1.90E-5& 7.13E-7& -9.97E-6& -7.02E-5&     -4.77E-4& -0.00101& -0.00424& -0.00693\\    
7.0& 3.52E-5& 2.99E-5& 1.25E-5& 3.75E-7& -6.82E-6& -4.79E-5&     -3.27E-4& -7.56E-4& -0.00297& -0.00491\\    
8.0& 2.48E-5& 2.10E-5& 8.74E-6& 2.16E-7& -4.89E-6& -3.43E-5&     -2.35E-4& -5.46E-4& -0.00217& -0.00362\\    
9.0& 1.82E-5& 1.54E-5& 6.40E-6& 1.33E-7& -3.65E-6& -2.55E-5&     -1.76E-4& -4.09E-4& -0.00164& -0.00275\\    
10.0& 1.39E-5& 1.17E-5& 4.85E-6& 8.63E-8& -2.80E-6& -1.96E-5&     -1.35E-4& -3.15E-4& -0.00127& -0.00214\\
\end{tabular}
\label{DotEHOverDotEtot.tbl}
\end{table}

\begin{table}
\caption{$\cal N$ (the relativistic correction to $\Omega^2/\dot\Omega =
d\Phi/d\ln\Omega$, the number of radians of orbital inspiral per unit 
fractional change of orbital angular velocity), as a function of orbital
radius $r/r_{\rm isco}$ and black-hole spin parameter $a$; cf.\ caption of
Table \protect\ref{Omega.tbl}.  
This table is accurate to four significant digits, and it
was computed using Eq.\ (\protect\ref{Nr}) and $\dot{\cal E}$ from
Table \protect\ref{ScrDotE.tbl}. 
Near the isco, ${\cal N} \protect\propto \tilde r - \tilde r_{\rm isco}
\protect\propto \tilde\Omega - \tilde\Omega_{\rm isco}$.
}
\begin{tabular}{d|dddddddddd}
$r/r_{\rm isco}$  &     $-$0.99 &     $-$0.9 &     $-$0.5 &      0.0 &      0.2 &      0.5 &      0.8 &      0.9 &      0.99 &      0.999 \\\tableline
1.000& 0.0& 0.0& 0.& 0.0& 0.0& 0.0&     0.0& 0.0& 0.0& 0.0 \\
1.001& 0.001966& 0.001995& 0.002150& 0.002471& 0.002685& 0.003262&     0.005188& 0.007911& 0.03914& 0.1960 \\
1.002& 0.003927& 0.003984& 0.004294& 0.004932& 0.005360& 0.006510&     0.01034& 0.01575& 0.07756& 0.3844 \\
1.005& 0.009777& 0.009917& 0.01068& 0.01227& 0.01332& 0.01616&     0.02560& 0.03888& 0.1886& 0.9077 \\
1.01& 0.01942& 0.01969& 0.02120& 0.02432& 0.02639& 0.03195&     0.05037& 0.07612& 0.3605& 1.656 \\
1.02& 0.03831& 0.03883& 0.04176& 0.04778& 0.05177& 0.06246&     0.09754& 0.1460& 0.6605& 2.785 \\
1.05& 0.09192& 0.09314& 0.09979& 0.1134& 0.1224& 0.1462& 0.2224&     0.3244& 1.296& 4.447 \\
1.1& 0.1721& 0.1742& 0.1857& 0.2090&     0.2242& 0.2639& 0.3865& 0.5431& 1.836& 4.964 \\
1.2& 0.3043& 0.3076& 0.3251& 0.3599& 0.3824& 0.4397& 0.6066&     0.8039& 2.133& 4.385 \\
1.3& 0.4078& 0.4116& 0.4321& 0.4725&     0.4983& 0.5628& 0.7424& 0.9427& 2.115& 3.725 \\
1.4& 0.4901& 0.4941& 0.5160& 0.5589& 0.5857& 0.6524& 0.8315&     1.022& 2.026& 3.228 \\
1.7& 0.6564& 0.6603& 0.6820& 0.7239&     0.7498& 0.8119& 0.9671& 1.117& 1.761& 2.373 \\
2.0& 0.7537& 0.7574& 0.7771& 0.8147& 0.8378& 0.8923& 1.022& 1.140&     1.589& 1.965 \\
2.5& 0.8452& 0.8482& 0.8644& 0.8955& 0.9142&     0.9579& 1.057& 1.142& 1.427& 1.637 \\    
3.0& 0.8949& 0.8975& 0.9110& 0.9370& 0.9526& 0.9886& 1.068& 1.134&     1.338& 1.477 \\
4.0& 0.9445& 0.9463& 0.9564& 0.9757& 0.9873&     1.014& 1.071& 1.116& 1.245& 1.324 \\   
5.0& 0.9673& 0.9688& 0.9768& 0.9917& 1.001& 1.022& 1.067& 1.101&     1.195& 1.250 \\
6.0& 0.9797& 0.9809& 0.9874& 1.001& 1.008& 1.025&     1.061& 1.089& 1.164& 1.207 \\
7.0& 0.9870& 0.9880& 0.9936& 1.005& 1.011& 1.025& 1.056& 1.080&     1.143& 1.177 \\
8.0& 0.9916& 0.9925& 0.9973& 1.005& 1.012& 1.025&     1.052& 1.072& 1.126& 1.156 \\
9.0& 0.9947& 0.9954& 0.9998& 1.008& 1.013& 1.024& 1.048& 1.066&     1.114& 1.140 \\
10.0& 0.9968& 0.9975& 1.001& 1.011& 1.013& 1.023&     1.045& 1.061& 1.104& 1.127 \\
\end{tabular}
\label{ScrN.tbl}
\end{table}

\newpage
\null
\newpage

\begin{table}
\caption{${\cal T}$ (the relativistic correction to $T$,
the time remaining until the
isco is reached) as a function of the orbital radius $r/r_{\rm isco}$ and
black-hole spin parameter $a$; cf.\ caption of Table \protect\ref{Omega.tbl}.
Near the isco, ${\cal T} \simeq (8/5) {\cal N}_{\rm orb}
\protect\propto (\tilde r - \tilde r_{\rm isco})^2
\protect\propto (\tilde\Omega - \tilde\Omega_{\rm isco})^2$.
We think this table is accurate to about 1 part in 500, except at
$r/r_{\rm isco} \protect\alt 1.2$ 
where the accuracy is about 1 part in 100.
The table was computed from 
Eqs.~(\protect\ref{rOmega}), (\protect\ref{Nr}), 
and (\protect\ref{calT}), using a cubic interpolation to 
$\dot{\cal
E}(r/r_{\rm isco})$ as given in Table \protect\ref{ScrDotE.tbl}.   
}
\begin{tabular}{d|dddddddddd}
$r/r_{\rm isco}$  &     $-$0.99 &     $-$0.9 &     $-$0.5 &      0.0 &      0.2 &      0.5 &      0.8 &      0.9 &      0.99 &      0.999 \\\tableline
1.000& 0.0& 0.0& 0.0& 0.0& 0.0& 0.0& 0.0& 0.0& 0.0& 0.0\\    
1.001& 4.08E-6& 4.13E-6& 4.40E-6& 4.93E-6& 5.28E-6& 6.16E-6& 8.93E-6&     1.26E-5& 5.01E-5& 2.22E-4\\ 
1.002& 1.63E-5& 1.65E-5& 1.75E-5& 1.97E-5&     2.10E-5& 2.46E-5& 3.56E-5& 5.02E-5& 1.99E-4& 8.75E-4\\    
1.005& 1.01E-4& 1.02E-4& 1.09E-4& 1.22E-4& 1.30E-4& 1.52E-4& 2.20E-4& 3.10E-4&     0.00122& 0.00525\\ 
1.01& 3.97E-4& 4.02E-4& 4.28E-4& 4.80E-4& 5.13E-4&     5.98E-4& 8.62E-4& 0.00121& 0.00470& 0.0196\\    
1.02& 0.00154& 0.00156& 0.00166& 0.00186& 0.00199& 0.00231&     0.00332& 0.00464& 0.0175& 0.0692\\    
1.05& 0.00883& 0.00893& 0.00949& 0.0106& 0.0113& 0.0131& 0.0186&     0.0256& 0.0897& 0.309\\ 
1.1& 0.0307& 0.0311& 0.0329& 0.0366&     0.0389& 0.0448& 0.0623& 0.0842& 0.266& 0.778\\    
1.2& 0.0950& 0.0960& 0.101& 0.112& 0.118& 0.134& 0.181& 0.237&     0.642& 1.53\\ 
1.3& 0.169& 0.171& 0.180& 0.196& 0.207& 0.233&     0.306& 0.390& 0.947& 2.00\\    
1.4& 0.243& 0.245& 0.257& 0.279& 0.293& 0.327& 0.420& 0.524&     1.17& 2.26\\ 
1.7& 0.432& 0.435& 0.452& 0.484& 0.503& 0.550&     0.674& 0.802& 1.49& 2.45\\    
2.0& 0.568& 0.571& 0.589& 0.624& 0.645& 0.696& 0.823& 0.949&     1.56& 2.33\\ 
2.5& 0.712& 0.715& 0.732& 0.766& 0.786& 0.834&     0.950& 1.06& 1.52& 2.04\\    
3.0& 0.796& 0.799& 0.815& 0.846& 0.864& 0.907& 1.01& 1.10& 1.45&     1.81\\ 
4.0& 0.886& 0.888& 0.901& 0.925& 0.940& 0.974& 1.05& 1.11&     1.34& 1.53\\ 
5.0& 0.929& 0.931& 0.941& 0.961& 0.973& 1.00& 1.06&     1.11& 1.27& 1.39\\ 
6.0& 0.953& 0.954& 0.963& 0.980& 0.990& 1.01&     1.06& 1.10& 1.22& 1.31\\ 
7.0& 0.967& 0.968& 0.976& 0.990& 0.999&     1.02& 1.06& 1.09& 1.19& 1.26\\    
8.0& 0.976& 0.978& 0.984& 0.997& 1.00& 1.02& 1.06& 1.09& 1.17&     1.22\\ 
9.0& 0.983& 0.984& 0.990& 1.00& 1.01& 1.02& 1.05& 1.08&     1.15& 1.19\\ 
10.0& 0.987& 0.988& 0.993& 1.00& 1.01& 1.02& 1.05&     1.07& 1.14& 1.17\\
\end{tabular}
\label{ScrT.tbl}
\end{table}

\begin{table}
\caption{${\cal N}_{\rm orb}$ (the relativistic correction to $N_{\rm orb}$,
the number of
orbits remaining until the isco is reached) as a function of orbital radius
$r/r_{\rm isco}$ and black-hole spin parameter $a$; cf.\ caption of Table
\protect\ref{Omega.tbl}.
Near the isco, ${\cal T} \simeq (8/5) {\cal N}_{\rm orb}
\protect\propto (\tilde r - \tilde r_{\rm isco})^2
\protect\propto (\tilde\Omega - \tilde\Omega_{\rm isco})^2$.
We think this table is accurate to about 1 part in 500, except at
$r/r_{\rm isco} \protect\alt 1.2$
where the accuracy is about 1 part in 100.
The table was computed from
Eqs.~(\protect\ref{rOmega}), (\protect\ref{Nr}),
and (\protect\ref{calNorb}), using a cubic interpolation to
$\dot{\cal
E}(r/r_{\rm isco})$ as given in Table \protect\ref{ScrDotE.tbl}.
}
\begin{tabular}{d|dddddddddd}
$r/r_{\rm isco}$  &     $-$0.99 &     $-$0.9 &     $-$0.5 &      0.0 &      0.2 &      0.5 &      0.8 &      0.9 &      0.99 &      0.999 \\\tableline
1.000& 0.0& 0.0& 0.0& 0.0& 0.0& 0.0& 0.0& 0.0& 0.0& 0.0\\    
1.001& 2.55E-6& 2.58E-6& 2.75E-6& 3.09E-6& 3.30E-6& 3.85E-6& 5.58E-6&     7.88E-6& 3.13E-5& 1.39E-4\\ 
1.002& 1.02E-5& 1.03E-5& 1.10E-5& 1.23E-5&     1.32E-5& 1.54E-5& 2.23E-5& 3.14E-5& 1.25E-4& 5.47E-4\\    
1.005& 6.31E-5& 6.39E-5& 6.81E-5& 7.63E-5& 8.16E-5& 9.52E-5& 1.38E-4&     1.94E-4& 7.62E-4& 0.00328\\    
1.01& 2.49E-4& 2.52E-4& 2.69E-4& 3.01E-4& 3.22E-4& 3.75E-4& 5.41E-4&     7.61E-4& 0.00294& 0.0123\\    
1.02& 9.73E-4& 9.85E-4& 0.00105& 0.00117& 0.00125& 0.00146& 0.00209&     0.00293& 0.011& 0.0435\\ 
1.05& 0.00566& 0.00572& 0.00608&     0.00678& 0.00723& 0.00837& 0.0119& 0.0163& 0.057& 0.196\\    
1.1& 0.0201& 0.0204& 0.0216& 0.0239& 0.0254& 0.0293& 0.0406&     0.0547& 0.172& 0.502\\ 
1.2& 0.0648& 0.0655& 0.069& 0.0759&     0.0803& 0.0911& 0.122& 0.160& 0.430& 1.03\\    
1.3& 0.119& 0.121& 0.127& 0.138& 0.146& 0.164& 0.214& 0.272&     0.657& 1.39\\ 
1.4& 0.177& 0.178& 0.187& 0.203& 0.212& 0.237&     0.303& 0.377& 0.837& 1.63\\    
1.7& 0.336& 0.339& 0.351& 0.376& 0.391& 0.428& 0.523& 0.623&     1.16& 1.95\\ 
2.0& 0.463& 0.465& 0.481& 0.509& 0.527& 0.569&     0.674& 0.780& 1.30& 2.00\\ 
2.5& 0.611& 0.614& 0.630& 0.660& 0.679&     0.722& 0.827& 0.927& 1.38& 1.92\\    
3.0& 0.708& 0.711& 0.726& 0.755& 0.773& 0.814& 0.911& 1.00& 1.38&     1.80\\ 
4.0& 0.820& 0.822& 0.835& 0.861& 0.876& 0.911& 0.992& 1.06&     1.33& 1.60\\ 
5.0& 0.879& 0.881& 0.892& 0.914& 0.927& 0.957& 1.02&     1.08& 1.28& 1.47\\ 
6.0& 0.914& 0.916& 0.926& 0.944& 0.955&     0.981& 1.04& 1.09& 1.25& 1.39\\    
7.0& 0.936& 0.938& 0.946& 0.963& 0.973& 0.995& 1.04& 1.09& 1.22&     1.33\\ 
8.0& 0.951& 0.953& 0.960& 0.975& 0.984& 1.00& 1.05& 1.08&     1.19& 1.28\\ 
9.0& 0.962& 0.963& 0.970& 0.983& 0.991& 1.01& 1.05&     1.08& 1.18& 1.25\\ 
10.0& 0.970& 0.971& 0.977& 0.989& 0.996& 1.01&     1.05& 1.08& 1.16& 1.22\\
\end{tabular}
\label{ScrNorb.tbl}
\end{table}


\begin{references}
\bibitem{LisaFit}In the computations underlying Figs.\ \ref{fig:muminyr} and
\ref{fig:muminisco} we used the following analytic fit to Bender and Stebbins'
\protect\cite{mdt} LISA noise curve $h^{\rm SA}_{\rm SN5,1yr}(f)$,
in which $f$ is measured in Hz: 
$$\sqrt{4.85\times 10^{-57}/f^4+ 
8.38\times 10^{-47} + 4.85*10^{-43} f^2}\;.\nonumber
$$
\end{references}
\end{document}